\documentstyle[prc,aps,preprint]{revtex} 
\begin{document}
\draft
\def\pmb#1{\setbox0=\hbox{#1}%
     \kern-.025em\copy0\kern-\wd0
      \kern.05em\copy0\kern-\wd0
       \kern-.025em\raise.0433em\box0}
\def\btau{\pmb{$\tau$}}
\def\bsigma{\pmb{$\sigma$}}
\def\bdiamond{\pmb{$\diamond$}}
\def\bbdiamond{\pmb\bdiamond}
\def\hat{\widehat}
\def\omeg{\bmatdieci\char '41}
\date{\today}

\title{The Dynamical Dipole Mode in Dissipative Heavy Ion Collisions}

\author{ V.Baran$^{1,2}$, M.Cabibbo$^{1,3}$, M.Colonna$^{1}$, M.Di Toro$^1$,
 and N.Tsoneva$^4$}

\address{$^1$ Laboratorio Nazionale del Sud, Via S. Sofia 44,
I-95123 Catania, Italy\\ and University of Catania}
\address{$^2$ NIPNE, Bucharest, Romania}
\address{$^3$ CEA-SACLAY, France}
\address{$^4$ INRNE, Sofia, Bulgaria}
\maketitle

\begin{abstract}

We study the effect of a direct Giant Dipole Resonance 
($GDR$) excitation in intermediate dinuclear 
systems with exotic shape and charge distributions formed in charge 
asymmetric fusion entrance channels.
A related enhancement of the $GDR$ gamma yield in the 
evaporation cascade of the fused nucleus is expected. The dynamical origin 
of such $GDR$ extra strength will show up in a characteristic anisotropy of 
the dipole gamma-emission.
A fully microscopic analysis of the fusion dynamics is performed with 
quantitative predictions of the $GDR$ photon yield based on a dynamics-
statistics coupling model. In particular we focus our attention on the 
energy and mass dependence of the effect. 
 We suggest a series of new experiments, in particular
some optimal entrance channel conditions. We stress the importance of
using the new available radioactive beams.

\end{abstract}

%{\bf Keywords: } 

\section{Introduction }

The isovector giant dipole resonance represents a well established
collective motion of finite nuclei extensively studied since more than
fifty years \cite{wou96}. The dependence on the nuclear structure of 
the reference state on top of which the collective mode is built has 
already suggested the use of $GDR$ properties in order to study nuclei 
far from the ground state.
The time structure of the $GDR$ mode actually allows a possibility
of using it as a probe of nuclear systems very far from normal
conditions, see the nice reviews \cite{sno86,gaa92}. \\
We can roughly estimate an oscillation period of 
$\frac{2\pi\hbar}{E_{GDR}} \simeq
80-100 fm/c$ and a mean life-time $\hbar/\Gamma_{GDR} \simeq 50fm/c$.
 Since the spreading width
is the largest contribution to the total width $\Gamma_{GDR}$,
$50fm/c$ is roughly also the time needed to build up the
collective $GDR$ mode in a Compound Nucleus (C.N.). All these time scales
are relatively short and this makes the $GDR$ an ideal probe to study
nuclear systems under extreme conditions.\\
Charge equilibration takes place on time scales of few units
in $10^2fm/c$: therefore in charge asymmetric entrance channels we expect
to see a "direct dipole" collective excitation, with special dynamical 
features, which should give an extra yield to the photon spectrum
in the $GDR$ region. The idea is that we can form a fused
dinuclear system with the charge not yet equilibrated and
therefore with some extra dipole strength of non statistical
nature. In this sense
   the $"Dynamical~Dipole"$ is a direct dipole contribution present in
   dissipative collisions initiated in charge asymmetric entrance channels.
   This new pre-equilibrium collective mode has been predicted some years
   ago \cite{cho92,cho93} and recently clearly observed in fusion 
   \cite{fli96,cin98,amo198} and deep-inelastic \cite{tro99,pap99} reactions.
   Some experiments have been performed also at intermediate energies, 
where other 
   dynamical entrance channel features can be present \cite{amo98,tud99}.
   The appearance of this prompt dipole effect gives important information 
   on the charge equilibration dynamics in connection to the reaction 
   mechanism. 
   A fully microscopic analysis is performed with quantitative predictions
   of the $GDR$ photon yield based on a dynamics-statistics coupling model.
   Independent information on the damping of the dipole mode in very excited
   nuclear systems can be derived. Moreover the coupling between
   isovector dipole and large amplitude isoscalar monopole oscillations
   can be studied, of large interest also for the discussion on the
   fate of the $GDR$ in hot nuclei. Since the effect is strongly
dependent on fusion dynamics we study the beam energy variation
of this extra dipole strength and we suggest some optimal conditions for
the relative observation.

\section{ GDR in charge asymmetric fusion: the dynamical dipole }

Fusion processes in charge asymmetric entrance channels give a clear
example on how emitted $GDR$ photons carry information on a Compound 
Nucleus ($CN$), formed during the reaction dynamics, very far from
normal conditions.

In dissipative collisions energy and angular momentum are quickly
distributed among all single particle degrees of freedom while charge
equilibration takes place on larger time scales \cite{gre84}.
Therefore for charge asymmetric entrance channels at the time of $CN$
formation we can easily have some relic of a pre-equilibrium $GDR$
from dipole oscillations in the isospin transfer dynamics
\cite{cho93}. The result will be an enhanced $GDR$ photon emission with
special features due to the non-statistical nature of this extra contribution.

An enhancement of the $GDR$ gamma ray yield has been observed
experimentally in fusion reaction between heavy ions with different neutron
to proton ratios. Flibotte et al. \cite{fli96}, in collision between
$^{40}Ca$ $(N/Z=1)$ and $^{100}Mo$, $(N/Z=1.38)$, at $4AMeV$ beam energy,
producing a $CN$ at $70 MeV$
excitation energy, counted a number of emitted $GDR$ photons over the whole 
cascade 
$16 \%$ larger than in the case when the same $CN$ was 
formed with a symmetric
$N/Z$ combination.
Similar effect has been observed by Cinausero et al.\cite{cin98}:
an increase of $36 \%$ resulted when they compared $GDR$ photon 
yields from a $CN$
populated  at $110$ MeV excitation energy via the reactions $^{16}O$ $(N/Z=1)$
+ $^{98}Mo$ $(N/Z=1.33)$  and  $^{50}Ti$ $(N/Z=1.18)$ + $^{64}Ni$ $(N/Z=1.28)$
respectively.
%Some experiments have been performed also at intermediate energies
%\cite{amo98} \cite{tud99} and in Deep Inelastic Collisions \cite{tro99},
%where other dynamical effects may be present.

In fact, theoretically, an extra-yield of gamma rays was predicted
some years ago by Chomaz et al., \cite{cho93},  in a simple
model of a $GDR$ phonon gas interacting with the $CN$.
The essential quantity in this approach is the mean number of excited $GDR$
phonons at the time when a $CN$ is formed, $n_{GDR}^{(0)}$. The emission of
$GDR$ photons is enhanced if $n_{GDR}^{(0)}$ is greater than the value
corresponding to a statistical equilibrium between the $GDR$ oscillator 
and the $CN$ heat bath. 
The $N/Z$ asymmetry of colliding heavy ions will certainly
trigger in the early stage of the reaction the charge equilibration process.
Several experiments and realistic simulations have indicated that the
related dipole oscillation has
the characteristics of a $GDR$ type quantal collective motion \cite{gre84},
\cite{hof79}, \cite{bon81}, \cite{sur88}, \cite{sur89}, \cite{bar96}. 
Expressed in terms of the
Brink-Axel hypothesis \cite{bri55} this means that a $GDR$ may be built
not only on equilibrium states, of a warm nucleus for example,
but also on non-equilibrium states during the $CN$ formation phase.
$n_{GDR}^{(0)}$ represents just the number of phonons in this
mode at the time of the $CN$ formation, $t_{CN}$.

Although it was assumed that the quantity $n_{GDR}^{(0)}$ is intimately 
related to
the presence of pre-equilibrium effects, the way it is determined and
affected by the projectile-target $N/Z$ asymmetry and entrance-channel 
dynamics has not been investigated yet.
From the previous discussion we clearly distinguish three main phases
in the entrance channel dynamics: {\it I - The approaching phase}, 
with the two
partners still keeping their own response properties; {\it II - The 
dinuclear phase}, with relative collective response; {\it III- The $CN$ 
formation}.
We will call $t=0$ the starting time of the phase {\it II},
i.e. the onset of the main dissipation mechanisms, including the
charge equilibration. In this phase a pre-equilibrium dipole mode
is present in the composite system, damped with a spreading width 
$\Gamma^\downarrow \equiv \hbar \mu(t)$, where the time dependence
is expressing a non-equilibrium situation. The correspondent
number of phonons will show an exponential decrease: 
\begin{equation}
n(t)=n_{i} e^{- \int_{0}^{t} \mu(t) dt }
\label{nphon}
\end{equation}
where $n_{i}$ is the number of phonons at the moment when charge 
equilibration begins. Fusion experiments seem to
indicate that this will happen once the decision for fusion is taken
\cite{gre84}. This means that when the system passes
from the strong absorption configuration through the barrier the neck 
is large 
enough to allow for an isospin collective motion. Therefore we 
can assume  that
the configuration reached soon after, when the interdistance between the
center of mass of the two nuclei is around the sum of their radii,
represents the starting point of the collective pre-equilibrium
$GDR$ mode. It is important to stress that
at this moment we have already reached a quite noticeable density
overlap and therefore the considered configuration is not
corresponding to the naive picture of the two nuclei as two
touching rigid spheres. This will become clear from density contour plots
and collective behaviours we will show later in microscopic dynamical
simulations.

In order to estimate $n_{i}$ we consider a harmonic oscillator description
for the $GDR$:  
\begin{equation}
H_{GDR}(t)=\frac{\Pi^{2}(t)}{2M} + \frac{M \omega^{2}(t)}{2} X^{2}(t),
\label{gdrene}
\end{equation} 
where $M= \frac{ZN}{(Z+N)} m$ is the reduced mass of the neutron-proton 
relative motion,
$m = 935 MeV$ being the nucleon mass, $N=N_{1}+N_{2}$, ($Z=Z_{1}+Z_{2}$),
the total number of neutrons (protons). Here $\Pi$ denotes the relative
momentum: 
\begin{equation}
\Pi= \frac{NZ}{A}(\frac{P_{p}}{Z}-\frac{P_{n}}{N})
\end{equation}
with $P_{p}$ ($P_{n}$) the center of mass momentum for protons
(neutrons), while $X=R_{p}-R_{n}$ is the distance between 
the centers of mass of the two components.
The initial number of phonons is defined as:
\begin{equation}
n_{i} = \frac{H_{GDR}(0)}{\hbar \omega(0)} \approx
\frac{1}{2 \hbar^{2}} M (\hbar \omega(0)) X^{2}(0)
\label{ninit}
\end{equation}
We use the above approximation in order to get a closed form
directly related to the initial charge asymmetry since 
$X(0)$ is given by: 
\begin{equation}
X(0)= \frac{Z_{1} Z_{2}}{N Z}(\frac{N_{1}}{Z_{1}}-\frac{N_{2}}{Z_{2}})
(R_{1}+R_{2})
\end{equation}
In Eq.(\ref{ninit}) we are neglecting the kinetic part contribution. 
We expect this approximation 
not to work at higher incident beam energies. This contribution
will be thoroughly accounted for from dynamical simulations, 
as discussed later.

We can consider for the initial elongated
shape a phonon energy \cite{RS80}:
\begin{equation}
\hbar \omega(0) = \frac{R}{R_{1}+R_{2}} E_{GDR} 
\end{equation}
where $R=r_{0} A^{1/3}$, ($r_{0}=1.2$), is the equilibrium radius of 
the $CN$ and
$E_{GDR} \simeq 78 A^{-1/3}$ is the energy of a phonon built on $CN$,
corresponding to the centroid of the $GDR$ spectrum in medium-heavy nuclei. 

In conclusion
we can finally express the mean number of phonons at the time when 
a $CN$ is formed as:
\begin{eqnarray}
n_{GDR}^{(0)} = n(t_{CN}) \approx  
\frac{1}{2 \hbar^{2}} M R (R_{1}+R_{2}) 
 \frac{Z_{1}^2 Z_{2}^2}{N^2 Z^2}(\frac{N_{1}}{Z_{1}}-\frac{N_{2}}{Z_{2}})^{2}
 E_{GDR} \exp{(- \mu_{ave} t_{CN})}  \nonumber \\
=
1.4 \frac {(A_{1}^{1/3}+A_{2}^{1/3})}{A} 
 \frac{Z_{1}^2 Z_{2}^2}{NZ}(\frac{N_{1}}{Z_{1}}-\frac{N_{2}}{Z_{2}})^{2}
 \exp{(- \mu_{ave} t_{CN})} 
\label{nphotheo}
\end{eqnarray}
where $\mu_{ave}$ is an average value of the spreading width:
\begin{equation}
\mu_{ave}= \frac{1}{t_{CN}} \int_{0}^{t_{CN}} \mu(t) dt 
\end{equation} 
We observe that $n_{GDR}^{(0)}$ will depend critically on fusion
dynamics (through the time scale for Compound Nucleus formation) 
as well as on 
properties of the spreading width $\mu(t)$.

After the time $t_{CN}$ we have to switch to the approach introduced 
in \cite{cho93} for the 
$CN$ decay phase.
The number of $GDR$ phonons is still time-dependent since 
charge equilibration is still going on, but now we can have
some feeding from the $CN$ heat bath with a width $\Gamma_{feed}
 \equiv \hbar \lambda$. The result is \cite{cho93} 
(now the origin of time is $t_{CN}$):
\begin{equation}
n_{GDR}(t) = \frac{\lambda}{\mu} {\Big[}{1 - (1 - n_{GDR}^{(0)} \frac{\mu}
{\lambda} ) \exp{(- \mu t)}}{\Big]}
\label{ngdr}
\end{equation}
which asymptotically leads to the statistical value: 
\begin{equation}
\frac{\lambda}{\mu} = \frac{\rho (E^* - E_{GDR})}{\rho (E^*)}
 \approx \exp{(-E_{GDR}/T)},
\label{neq}
\end{equation}
where $\rho(E)$ is the level density, $E^*$ is the
$CN$ excitation energy and $T$ the corresponding temperature.
From this discussion we see that the quantity Eq.(\ref{ngdr}) can
be directly used as the dipole strength in the evaluation
of $\gamma$-decay in the $CN$ evaporation cascade. From the
detailed balance we have a statistical $\gamma$ emission rate:
%for each dipole component,
\begin{equation}
R_{\gamma}(E_{\gamma}) = \frac{\rho (E^* - E_{\gamma})}{\rho (E^*)}
 \frac{\sigma_{abs}(E_{\gamma})}{3} \frac{E_{\gamma}^2}{(\pi \hbar c)^2}
= \\
\frac{\rho (E^* - E_{GDR})}{\rho (E^*)}
{\Big[}  \frac{\rho (E^* - E_{\gamma})}{\rho (E^* - E_{GDR})}
\frac{\sigma_{abs}(E_{\gamma})}{3} \frac{E_{\gamma}^2}{(\pi \hbar c)^2}
{\Big]}
\label{rgam}
\end{equation}
where $\sigma_{abs}(E_{\gamma})$ is the absorption $\gamma$-dipole
cross section.
This leads, in presence of a pre-equilibrium 
contribution,
to a time dependent quantity:
\begin{equation}
R_{\gamma}(E_{\gamma},t) = n_{GDR}(t) 
{\Big[}  \frac{\rho (E^* - E_{\gamma})}{\rho (E^* - E_{GDR})}
\frac{\sigma_{abs}(E_{\gamma})}{3} \frac{E_{\gamma}^2}{(\pi \hbar c)^2}
{\Big]}
\label{rgamt}
\end{equation}
with $n_{GDR}(t)$ given by Eq.(\ref{ngdr}).

The final yield of photons emitted can be then evaluated using a 
time-dependent evaporation cascade procedure, see ref.\cite{Maur2}.
From Eq.s(\ref{rgamt}) and (\ref{ngdr}), once  the $CN$ 
initial excitation energy
(temperature) is fixed, the results will be critically dependent on
$n_{GDR}^{(0)}$ and $\Gamma^\downarrow (T)$.

To show how the method is working and the sensitivity to the above
parameters in Fig.1 we report some results for the systems studied
by Flibotte et al \cite{fli96}. The curves represent the 
"subtracted" photon spectra (i.e.normalized to a statistical
emission without $GDR$): in $(a)$ the data for the two systems,
$Ca+Mo$ (charge asymmetric, dashed) and $S+Pd$ (more symmetric, solid),
and in $(b),(c),(d)$ the results obtained from $10^5$ runs of a
Monte Carlo evaporation cascade \cite{Maur1,Maur2} 
with the same initial conditions for
the $CN$ and varying only
$n_{GDR}^{(0)}$ and $\Gamma^\downarrow (T)$.
The best results (Fig.1b) are obtained with the values
$n_{GDR}^{(0)} = 0.14$ and $\Gamma^\downarrow = 8 MeV$. Now while it is
reasonable to have a $GDR$ damping width larger than in the
ground state of the same compound nucleus $^{140}Sm$ 
($\Gamma^\downarrow = 4.8 MeV$
used in Fig.1c) the choice of $n_{GDR}^{(0)}$ requires a deeper
dynamical justification. This will be discussed later in a detailed
connection to the fusion dynamics that will allow some
interesting predictions on the beam energy dependence of this
$GDR$ extra strength.

\section{Dynamical Dipole Signature on Angular Distributions}

We note that the pre-equilibrium contribution has some special 
dynamical features, being associated with a dipole component on the
reaction plane, actually along the symmetry axis of the
dinuclear system. If we call this as the $z-$component, we expect
no dipole strength present on the $x,y$ components at the time
of the $CN$ formation. The corresponding number of phonons will
also reach asymptotically the statistical value Eq.(\ref{neq})
(times $1/3$ on average) but starting from zero at $t=t_{CN}$
\cite{bri90}.

We have then 
time dependent $\gamma$-dipole emission rates
 different for each
component in the intrinsic reference system of the dinucleus:
\begin{eqnarray}
R_{\gamma,z}(E_{\gamma},t) = 
{\Big[} \frac{\lambda}{3 \mu}(1 - \exp{(- \mu t)}) + {n_{GDR}^{(0)}
 \exp{(- \mu t)}} {\Big]} \frac{\sigma_{abs}(E_{\gamma})}{3}
 \frac{E_{\gamma}^2}{(\pi \hbar c)^2} \nonumber \\
 \equiv W_z(t)
\frac{\sigma_{abs}(E_{\gamma})}{3}
 \frac{E_{\gamma}^2}{(\pi \hbar c)^2}
\label{rgamz}
\end{eqnarray}
\begin{eqnarray}
R_{\gamma,x}(E_{\gamma},t) = R_{\gamma,y}(E_{\gamma},t) = 
{\Big[} \frac{\lambda}{3 \mu}(1 - \exp{(- \mu t)}) {\Big]} 
\frac{\sigma_{abs}(E_{\gamma})}{3}
 \frac{E_{\gamma}^2}{(\pi \hbar c)^2} \nonumber \\
\equiv W_{x,y}(t)
\frac{\sigma_{abs}(E_{\gamma})}{3}
 \frac{E_{\gamma}^2}{(\pi \hbar c)^2}
\label{rgamxy}
\end{eqnarray}
with $n_{GDR}^{(0)}$ given by Eq.(\ref{nphotheo}).

We remark the physical meaning of the time dependent weights
$W_{z,x,y}(t)$ defined before, they represent the number
of dipole phonons on each component at each time step.
In Fig.2 we report a typical behaviour of the weights.
We have chosen the parameters corresponding to the $Ca+Mo$
system of ref.\cite{fli96}, see before.

In particular we note the extreme values at $t=0$ ($CN$ formation time)
and asymptotically $t \rightarrow \infty$ (full charge equilibration):
\begin{eqnarray}
W_z (0) = n_{GDR}^{(0)}~~,~~ W_z (\infty) = \frac{\lambda}
{3 \mu} \\
W_{x,y} (0) = 0 ~~,~~W_{x,y} (\infty) = \frac{\lambda}{3 \mu},
\end{eqnarray}
i.e. the $GDR$ phonons are initially only on the $z-$component
and finally uniformly distributed on the three axes. This property
will have important observable effects on the 
$\gamma-$angular distributions.

The angular distribution of the emitted photons can be written as
\cite{thir88,gaa92,bra95,bagro} :
\begin{equation}
W(\theta, E_{\gamma}) = W_0 (1 + a_2(E_{\gamma}) P_2(\theta))
\end{equation}
where $P_2(\theta)$ is a Legendre Polynomial in the polar angle $\theta$
between the direction of the emitted $\gamma-$ray and the beam axis.
With our choice of the oscillation axes we have $(x,z)-$stretched
and $y-$unstretched transitions \cite{thir88} 
and a time-dependent anisotropy
parameter:
\begin{equation}
a_2(E_{\gamma}, t) = \frac{W_y/2 - (W_z+W_x)/4}{W_x+W_y+W_z},
\end{equation}
with the time dependent weights $W_{z,x,y,}(t)$ given
by Eq.s (\ref{rgamz},\ref{rgamxy}).
The $z-$ extra strength contribution should therefore lead to
clear negative values around the $GDR$ energy
{\it also in absence of an intrinsic deformation of the fused nucleus}.
We remark that all the previous equations can be easily
extended to the use of different Lorentzians for the various axis
contributions to $\sigma_{abs}$, to take into account the presence
of deformations. The dynamical dipole effect will enhance the
corresponding characteristic anisotropies in the $\gamma-$
angular correlations.

\section{The interplay between fusion dynamics and pre-equilibrium GDR}

In this section we present a fully microscopic analysis of the fusion
reaction for some systems of interest, to look in particular
at the pre-equilibrium GDR population.
The calculations are performed in the framework of the BNV transport equation
which incorporates in a self-consistent way the mean field and two-body
collisions dynamics \cite{bonbnv}. The numerical accuracy has been
largely improved in order to have a good description also of low
energy fusion reactions, see discussion in \cite{bar96,bagro}. 
Moreover in order to further reduce numerical fluctuations the
results presented here are obtained from an average over several
events.
For the mean field we have adopted a Skyrme-like parametrization,
 as described in \cite{bar96}, which well reproduces Nuclear Matter
saturation properties (equilibrium density,
binding energy, compressibility and symmetry energy). In the collision integral
we use in medium reduced nucleon-nucleon cross sections,
 isospin as well as energy and angular dependent \cite{LiMach}.

We have focussed our attention on the fusion reaction induced by
$^{16}O$ $(N/Z=1)$ on  $^{98}Mo$ $(N/Z=1.33)$ (see \cite{cin98}).  
We have studied in a relative large range of incident beam energy, 
at $4 MeV/n$,
$8 MeV/n$, $14 MeV/n$ and $20 MeV/n$, the evolution of the composite 
system along the fusion path. 
We will consider results that correspond to an impact parameter
$b=0$ as well representative of the features we are looking for.

As discussed in the previous section an essential point is
to evaluate the starting time of phase $II$, when charge
equilibration and the relative collective dipole are present in the
dinuclear system. This time can be directly extracted from
the collision simulation just comparing the evolution
of the distance between Projectile-Target Centers of Mass, $d_{PT}(t)$,
 and of the distance between Neutron-Proton Centers of Mass, $X(t)$,
 reported in Fig.3. The two quantities are roughly proportional
up to the time of the onset of a collective dipole response of the
dinuclear system. At this time step, chosen as $t=0$ in Fig.3,
the quantity $X(t)$ clearly shows an acceleration due to 
the symmetry term of the dinuclear mean field. We remark that
for all energies the value of the $d_{PT}$ distance corresponding
to the bending of the "dipole" variable $X(t)$ is very close
to the sum of the two radii, in agreement with the discussion
of the previous section.

In all the following plots we will choose as $t=0$ such
starting time of the dinuclear dynamics, roughly corresponding to the
"touching" configuration. 

At each time step we can evaluate the dinuclear dipole moment in 
coordinate and momentum space:
\begin{eqnarray}
DR(t)&=&\frac{NZ}{A} X(t) \\
DK(t)&=&\frac{\Pi(t)}{\hbar}
\end{eqnarray}
The results are shown in Fig.4. The out of phase behaviour of the two dipoles,
clear signature of a collective dinuclear response, is indeed
starting just
after $t=0$ (touching configuration) practically for all energies.
We remark a smaller initial amplitude of the oscillation at the lowest
($4AMeV$) and highest ($20AMeV$) energies, as well as an increase
of the damping with the beam energy. In conclusion it clearly 
appears an optimum range of energies for the observation of the
dynamical dipole effect, well above the Coulomb threshold
(about $4AMeV$ in the $O+Mo$ system) but also well below the
Fermi energy domain. This point will be further analysed in the following.

Fig.5 shows phase-space trajectories of the $GDR$, i.e. the time
evolution of the $DK-DR$ correlation. The spiral curves are nicely
revealing an out of phase behaviour in presence of some damping.
The spiralling trend is starting at the touching configuration
($t=0$, left points of the curves), maybe with some delay in the 
lowest energy case. The centre is reached when charge
equilibration is achieved. From this time onwards the $GDR$ mode
will be only of statistical nature.
The spiral shows a faster collapse to the central region at high
beam energy (Fig.5d) since we have a larger damping of the 
dynamical collective
motion, as already seen in Fig.4.

The amplitude reduction at $4AMeV$ and $20AMeV$ is also evident: the
nature of the effect is however different in the two cases. At
low energy, just above the Coulomb barrier, the neck is formed
quite slowly (see Fig.6) and so the collective dipole response of
the dinuclear system is actually starting sometime after the
touching configuration, when some charge equilibration has
already taken place. At $20AMeV$ the fusion dynamics is fast
but now we can have some prompt particle emission that will
reduce the dipole moment of the dinucleus. Just by chance we see
 a quite good spiralling behaviour at $8AMeV$, where the experiment 
has 
been performed by Cinausero et al. \cite{cin98} with a very
clear evidence of the dynamical dipole enhancement. 

Now we can dynamically evaluate the number of pre-equilibrium 
$GDR$ phonons $n_{GDR}^{(0)}$ present at the time of the $CN$
formation: as shown in the previous section this quantity is
essentially ruling the extra emission of $GDR$ $\gamma$-rays.
At each time step the average number of phonons $n(t)$
can be determined from the simulations just dividing
the total oscillator energy, as defined by Eq.(\ref{gdrene}),
by the one phonon energy at that
instant, $\hbar \omega(t)$, i.e. corresponding to the
deformation at the time $t$. The results are presented in Fig.7.
The arrows show the initial value $n(0)$ predicted from
Eq.(\ref{ninit}), i.e. without dynamical effects.
The dashed curve in all four pictures is the analytical 
expectation of
Eq.(\ref{nphon}) 
using a constant spreading width value, 
$\Gamma^\downarrow=5MeV$. 
We remark a quite good overall agreement at low beam energies 
$4$ and $8 MeV/n$, $(a)$ and $(b)$. With increasing energy we clearly see 
in the number of phonons obtained from the simulations a 
 faster damping which quickly compensates the higher
starting point. This is particularly evident at
$20 MeV/n$ (Fig.7d) where we have to consider a 
quite larger value for the spreading
width joint to a sizeable kinetic term contribution to 
the initial number of phonons $n(t=0)$.

This time evolution of the number of dipole phonons in the 
dinuclear system will strongly influence the
quantity we are looking for, $n_{GDR}^{(0)}$, given by the value
of $n(t)$ at the time of Compound Nucleus formation $t_{CN}$,
 Eq.(\ref{nphotheo}), the latter being fixed from equilibrium conditions
in coordinate and momentum space.
In Fig.8
 we plot the time
evolution of the mass quadrupole moment in $\vec r$-space at the four incident 
energies.
Generally we remark two distinct behaviours in the evolution: a faster
decay until the first minimum is reached, followed by a smoother trend
accompanied by small oscillations. During the first stage the
conversion of relative motion into heat takes place and therefore, in the limits
of some more time needed for a final thermal equilibration, 
the moment when the first
minimum is attained will provide us the Compound Nucleus formation time.
Our evaluation is in agreement with the information obtained by looking at
the quadrupole in $\vec p$-space (see Fig.9), which is a good probe
of the energy relaxation time. We can extract $t_{CN}$ values
of $120~fm/c$ at $4AMeV$, $80~fm/c$ at $8AMeV$, and around $50~fm/c$ at
the higher energies $14AMeV$ and $20AMeV$. From Fig.7 we
get the corresponding $n_{GDR}^{(0)} = n(t_{CN})$ parameters:
$0.01$ at $4AMeV$ going up to $0.07$ at $8AMeV$ and
 finally reducing again down to $0.05$ at
$14AMeV$ and to $0.04$ at $20AMeV$.

To test the correctness of our procedure we have performed
a complete evaporation cascade calculation using the above
extra dipole strength for the system $O+Mo$ at $8AMeV$.
The "subtracted" spectra 
(see definition in section II) are shown in Fig.10, from
$2*10^5$ Montecarlo events, compared with experimental
data from ref.\cite{cin98}. The agreement is quite good.

Since the time scale of fusion dynamics is playing an
important role on the amount of extra dipole strength
present at the time of $CN$ formation we expect to see
some dependence of the effect also on the mass symmetry
in the entrance channel, which is strongly affecting
the fusion time, see \cite{Maur2} and ref.s therein.
In order to check this point we have studied the system
$^{50}Cr~(N/Z=1.08)~+~^{64}Ni~(N/Z=1.28)$ with almost
the same charge asymmetry of $O+Mo$ but much more
symmetric in mass. We have considered central collisions at 
the beam energies
$3.5~and~5.5AMeV$ corresponding to the same c.m energies above
the coulomb barrier of the first two cases of the
$O+Mo$ system. The relative fusion paths are shown in Fig.11
for $100fm/c$ starting from the touching configuration,
to be compared with the equivalent Fig.6 of the $O+Mo$.
At both energies it is quite evident the longer 
shape equilibration time. 
In Fig.12a,b (to compare with Fig.3a,b for $O+Mo$)
we see how the dinuclear formation time ($t=0$ in our
convention) can be again well deduced from the simulations.
In Fig.12c,d we show the time evolution of the correlation between
momentum and space dipoles. If we compare with the analogous
Fig.5a,b of the mass asymmetric $O+Mo$ case we clearly 
observe a quite larger damping of the collective dipole mode
in the dinucleus, due to the larger excitation energy
available in the more symmetric system and the related increase
of the $GDR$ spreading width \cite{bagro,war98}.
The slower shape equilibration is quantitatively analysed
in Fig.13a,b (see the corresponding Fig.8a,b for $O+Mo$)
and finally the time evolution of the number of dipole 
phonons is presented in Fig.13c,d, to compare with the equivalent
Fig.7a,b for the mass asymmetric case. We remark that in spite
of the larger initial value $n(t=0)$, as expected from
Eq.(\ref{ninit}) since $R_1+R_2$ is larger in case of mass
symmetry, we finally get a much smaller value of
$n_{GDR}^{(0)} = n(t_{CN})$ for two main reasons,
the larger damping and the longer $CN$ formation time.

We have finally checked the method also for the ref.\cite{fli96}
system, $Ca+Mo$ at $4AMeV$, where the best results are
obtained with a $n_{GDR}^{(0)}=0.14$, see Fig.1.
In Fig.14 we present the time evolution of the relevant
quantities, mass quadrupole moment, dipole moments in 
$\vec r$- and $\vec p$- space and average number of phonons.
With respect to the $O+Mo$ case at the same energy we see 
a larger
initial dipole strength but with a stronger damping and a
slower fusion process. We have a $t_{CN} \simeq 120fm/c$
and a corresponding $n_{GDR}^{(0)}$ ranging between
$0.12$ and $0.15$, as needed in the fitting procedure
of Fig.1.

\section{ Summary and conclusions }

We have shown that important information 
on the early stage of the fusion path can be obtained studying 
charge-asymmetric reactions. In fact, in this case it is possible to reveal 
a "direct" dipole oscillations ({\it the dynamical dipole}), 
related to the charge equilibration 
dynamics, which leads to some extra $GDR$ strength in the statistical
decay of the fused system.

The dynamical nature of such pre-equilibrium contribution, i.e.
corresponding to dipole oscillations on the reaction plane, is expected to
show up in an anisotropic $\gamma$-emission, also in absence of deformations 
in the compound nucleus. 

In this work we have investigated how $N/Z$ asymmetry and fusion dynamics
are affecting the properties of the pre-equilibrium $GDR$ remnants.  
We have shown the existence of a $GDR$ collective mode,
during the formation phase of the compound nucleus, i.e. built
on nonequilibrium states. This has  allowed us to express the mean number 
of dipole phonons present in the fused nucleus in terms of the initial 
isospin asymmetry, $CN$ formation time and $GDR$ spreading width.
We can make quite accurate predictions on the optimal choice of the
reaction partners, in particular on the interplay between charge
and mass asymmetry in the entrance channel.

Particularly interesting is to follow
the energy dependence of the effect. We expect to see a
{\it rise and fall} of the dynamical dipole contribution.

At low energies, just above the Coulomb barrier, the effect is
strongly reduced for two combined reasons: 
\begin{itemize}
\item 
i) A delay in the dinucleus
formation (slow neck dynamics) and relative collective response.
The pre-equilibrium dipole oscillation of the composite system
will have an initial smaller amplitude since some charge equilibration
has already taken place.
\item 
ii) A longer Compound Nucleus formation time is decreasing
the average number of phonons present in the fused system
due to the $GDR$ damping.
\end{itemize}

At higher energies, close to the Fermi energy domain, the effect
is again reduced on average for two main reasons:
\begin{itemize}
\item 
i) We have a relevant pre-equilibrium particle emission
(i.e. incomplete fusion events) with a direct reduction of the
initial charge asymmetry.
\item 
ii) In the fusion processes we have a large excitation energy
deposited in the composite system with a related increase of the
$GDR$ spreading width leading to a fast decrease of the number 
of dipole phonons.
\end{itemize}

Mass symmetry in the entrance channel is also strongly affecting
the dynamical dipole extra-strength. For the same charge asymmetry
and the same incident energy above the Coulomb barrier more
mass symmetric systems are expected to show a more reduced
pre-equilibrium $GDR$ strength for two main reasons,
a longer compound nucleus formation time and a larger
$GDR$ spreading width due to the larger excitation energy
that can be reached in the dinuclear configuration.
These two effects are clearly compensating the larger
value of the dipole moment at the touching time.
An increase of the beam energy will not help for the 
reasons listed above, pre-equilibrium and larger damping
of the collective mode. In conclusion mass-symmetric
entrance channels are strongly quenching the dynamical
dipole effect at all energies.

Moreover a new dynamical feature is appearing at beam energies
above $20AMeV$, some large monopole oscillations in the
entrance channel dynamics \cite{pia99,suo98,war98}.
The dynamical dipole strength is reduced and more fragmented, 
see the simulations in ref.\cite{war98}, although still present as
confirmed from very recent data \cite{amo98,tud99}.
The study of the dynamical dipole effect in the Fermi energy
domain, although experimentally quite difficult, would 
therefore bring new independent information on the spreading
width of hot $GDR$ and in particular on the coupling 
to an expanding collective mode

Finally, for charge asymmetric reactions,  
the enhanced dipole emission could be an
interesting cooling mechanism to favour the fusion
of very heavy nuclear systems.

From the interplay between charge and
shape equilibration time-scales we can also suggest new experiments to
study the dipole propagation in excited nuclei. 
The use of radioactive beams will enhance the
possibility of such observations.

\subsection*{Acknowledgements}
We warmly thank D.M.Brink and Ph.Chomaz for many pleasant
and enlightning discussions. We are deeply grateful to the TRASMA
experimental group of LNS-Catania, in particular to F.Amorini, G.Cardella, 
M.Papa and S.Tudisco for the availability of their "fresh" data and for a
stimulating interface.
Two of us (V.B. and N.T.) acknowledge the INFN support for their
stay at the Laboratori Nazionali del Sud, Catania. They also deeply
thank the LNS staff for the warm atmosphere and the good working 
conditions.

\newpage

\section*{Figure captions}

\begin{description}

\item[Fig. 1] Subtracted $\gamma$ spectra from a $^{140}Sm~CN$ formed
at $E^*=71MeV$ in the charge asymmetric ($Ca+Mo$ - dashed lines) and 
symmetric ($S+Pd$ - solid lines) entrance channel: a) Experiment
\cite{fli96}; b), c), d), simulations (see text) with 
($n_{GDR}^{(0)}, \Gamma ^\downarrow $) respectively equal to
($0.14, 8MeV$), ($0.6, 8MeV$), ($0.14, 4.8MeV$). $10^5$
Montecarlo events.

\item[Fig. 2]
Time evolution of the number of phonons on the different
intrinsic axes for a $^{140}Sm~CN$ formed
at $E^*=71MeV$ in the charge asymmetric $Ca+Mo$ 
entrance channel. The used parameters are $E_{GDR}=15MeV$, $T=2MeV$,
 $n_{GDR}^{(0)}=0.14$ and $\Gamma ^\downarrow = 8MeV$.

\item[Fig. 3]
Time evolution of the distance between Projectile-Target Centers
of Mass $d_{PT}(t)$ (solid lines, left scales in $fm$) and of the
distance between Neutron-Proton Centers of Mass $X(t)$ (dashed
lines, right scales in $fm$) for the $O+Mo$ system at beam energies: 
a) $4AMeV$, b) $8AMeV$, c) $14AMeV$, d) $20AMeV$. The $t=0$
choice is discussed in the text.

\item[Fig. 4] Time evolution of the dipole moment in $\vec r$-space, 
$DR$
(solid lines, left scales in $fm$) and in $\vec p$-space, $DK$ 
(dashed lines, right scales in $fm^{-1}$)
for the  $O+Mo$ reaction. $t=0$ corresponds to the touching
configuration (see text).
The labels correspond to the
same energies of Fig.3.

\item[Fig. 5] Phase space trajectory of the entrance channel
dipole for the $O+Mo$ reaction. The labels correspond to the
same energies of Fig.3.

\item[Fig. 6] Density plots of the neck dynamics for the $O+Mo$
system at the two energies $4AMeV$ and $8AMeV$.

\item[Fig. 7] Time evolution of the number of dipole phonons
for the $O+Mo$ reaction (solid lines). The labels correspond to the same
energies of Fig.3. The dashed line is the reference curve
discussed in the text, with a constant spreading width.

\item[Fig. 8] Time evolution of the mass quadrupole moment for
the $O+Mo$ reaction (in $a.u.$). The labels correspond to the same 
energies of Fig.3.

\item[Fig. 9] Time evolution of the quadrupole moment in momentum
space for the $O+Mo$ reaction (in $a.u.$). 
The labels correspond to the same energies of Fig.3.

\item[Fig. 10] Subtracted $\gamma$ spectra from a $^{114}Sn~CN$ formed
at $E^*=108MeV$ in the charge asymmetric ($O+Mo$ - dashed lines) and 
symmetric ($Ti+Ni$ - solid lines) entrance channel: a) Experiment
\cite{cin98}; b) simulations (see text) with 
$(n_{GDR}^{(0)}, \Gamma ^\downarrow)$ respectively equal to
($0.07, 7MeV$). $2*10^5$ Montecarlo events.

\item[Fig. 11] Density plots of the neck dynamics for the $Cr+Ni$
system at the two energies $3.5AMeV$ and $5.5AMeV$.

\item[Fig. 12]
Collision of the more mass symmetric $Cr+Ni$ system at the two
beam energies $3.5~and~5.5AMeV$.
(a) and (b): quantities like in Fig.3.
(c) and (d): quantities like in Fig.5.

\item[Fig. 13]
Collision of the more mass symmetric $Cr+Ni$ system at the two
beam energies $3.5~and~5.5AMeV$.
(a) and (b): quantities like in Fig.8.
(c) and (d): quantities like in Fig.7.

\item[Fig. 14] Time evolution of mass quadrupole moment (a), 
$\vec r$-dipole moment (b), $\vec p$-dipole moment (c) and average number
of dipole phonons (d) for the reaction $Ca+Mo$ at $4AMeV$.

\end{description}


\begin{references}
\bibitem{wou96} A. Van der Woude, Nucl.Phys. A599 (1996) 393c
\bibitem{sno86} K.A.Snover, Annu.Rev.Nucl.Part.Sci. 36 (1986) 545
\bibitem{gaa92} J.J.Gaardhoje, Annu.Rev.Nucl.Part.Sci. 42 (1992) 483
\bibitem{cho92} Ph. Chomaz in "Proceedings of the 6th Franco-Japanese
Colloquium", St. Malo, France, 1992, edited by N. Alamanos, S. Fortier and
F. Dykstra, Saclay (1993)
\bibitem{cho93} Ph.Chomaz, M.Di Toro and A.Smerzi, Nucl.Phys. 
        A563 (1993) 509
\bibitem{fli96} S.Flibotte et al., Phys.Rev.Lett. 77 (1996) 1448
\bibitem{cin98} M.Cinausero et al., Nuovo Cimento A111 (1998) 613
\bibitem{amo198} F.Amorini et al., Phys.Rev. C58 (1998) 987
\bibitem{tro99} M.Trotta et al., RIKEN Review 23 (1999) 96.\\
  M.Sandoli et al., Eur.Phys.J. A6 (1999) 275
\bibitem{pap99} M.Papa et al. , Eur.Phys.J. A4 (1999) 69.
\bibitem{amo98} F.Amorini, TRASMA exp., Ph.D.Thesis LNS Catania 1998.
\bibitem{tud99} S.Tudisco, TRASMA exp., Ph.D.Thesis LNS Catania 1999.
\bibitem{gre84} C. Gregoire, in "Proceedings of the Winter College
on Fundamental Nuclear Physics", vol. 1, Trieste, Italy, 1984,
edited by K. Dietrich, M. Di Toro, H. J. Mang, pp.497-643.
\bibitem{hof79} H. Hofmann et al., Z. Physik A293 (1979) 229.
\bibitem{bon81} P. Bonche, N. Ngo, Phys. Lett. B105  (1981) 17.
\bibitem{sur88} E. Suraud, M. Pi, P. Schuck, Nucl. Phys. A482
 (1988) 187c.
\bibitem{sur89} E. Suraud, M. Pi, P. Schuck, Nucl. Phys. A492
 (1989) 294.
\bibitem{bar96} V. Baran et al., Nucl. Phys. A600 (1996) 111.  
\bibitem{bri55} D.M.Brink, D.Phil.Thesis, Oxford 1955.\\
          P.Axel, Phys.Rev.126 (1962) 671
\bibitem{RS80} P.Ring and P.Schuck, {\it The Nuclear Many-Body Problem},
 Springer-Verlag, New York 1980
\bibitem{Maur2} M. Cabibbo, V. Baran, M. Colonna and M. Di Toro,
Nucl. Phys. A637, (1998) 374.
\bibitem{Maur1} M. Cabibbo, $MONTECASCA~Code$, Ph.D. Thesis, 
Univ. of Catania 1998
%\bibitem{Morsch} H.P. Morsch et al., Phys. Rev. Lett. 64,
%(1990) 1999.
\bibitem{bri90} D.M.Brink, Nucl.Phys. A519 (1990) 3c
\bibitem{thir88} P.Thirolf et al., Nucl.Phys.A482 (1988) 93c
\bibitem{bra95} A.Bracco et al., Phys.Rev.Lett. 74 (1995) 3748
\bibitem{bagro} V.Baran et al., Nucl.Phys. A599 (1996) 29c,\\
      V.Baran et al., Progr.Part.Nucl.Phys. 38 (1997) 263
\bibitem{bonbnv} A.Bonasera et al., Phys.Lett. B221 (1989) 233.\\
	A.Bonasera et al., Phys.Rep. 243 (1994) 1
\bibitem{LiMach} G.Q. Li and R. Machleidt,
 Phys. Rev. C48  (1993) 1702; Phys. Rev. 
 C49 (1994) 566;\\
 G.Q. Li, private communications.
%\bibitem{SRLS} A.Schnell, G.R\"opke, U.Lombardo and H.J.Schulze,
 Phys. Rev. {\bf C57} 503 (1998)
%\bibitem{Bon90} A. Bonasera, M. Di Toro and F. Gulminelli,
%Phys. Rev. {\bf C42}, 966 (1990).
\bibitem{pia99} P.Piattelli et al., Nucl.Phys.A649 (1999) 181c
\bibitem{suo98} T.Suomijarvi, RIKEN Symp.on Dynamics in hot nuclei, p.1,1998
\bibitem{war98} M.DiToro et al., Acta Phys.Polonica B30 (1999) 
    1331-1352
\end{references}
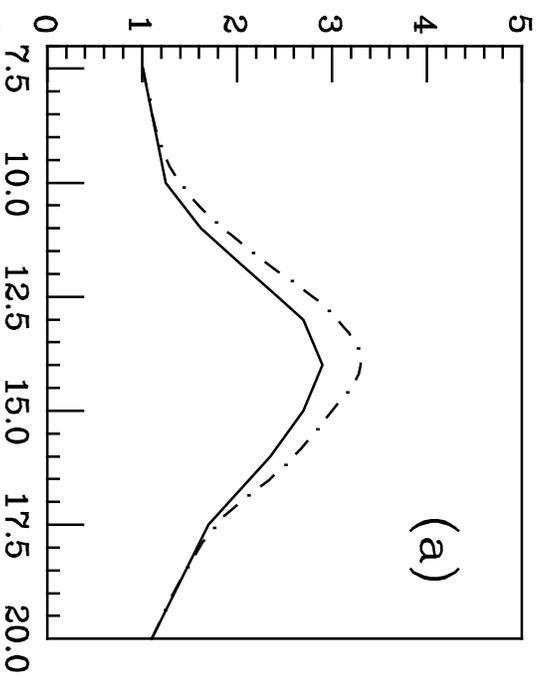
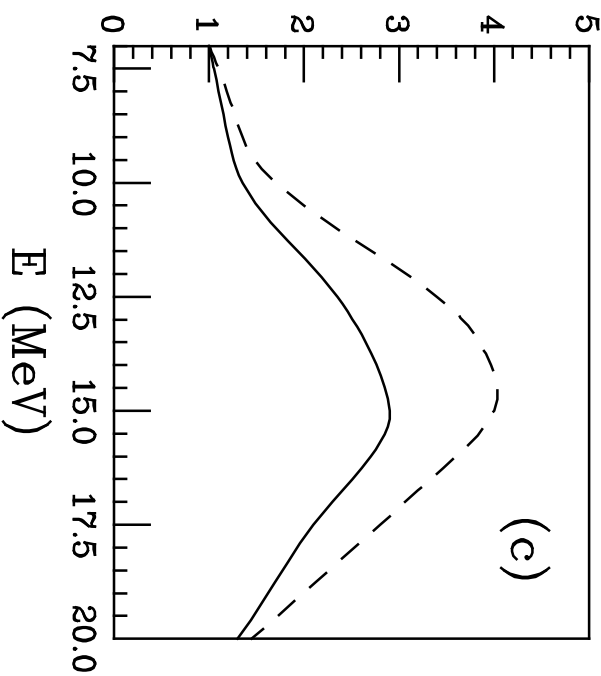
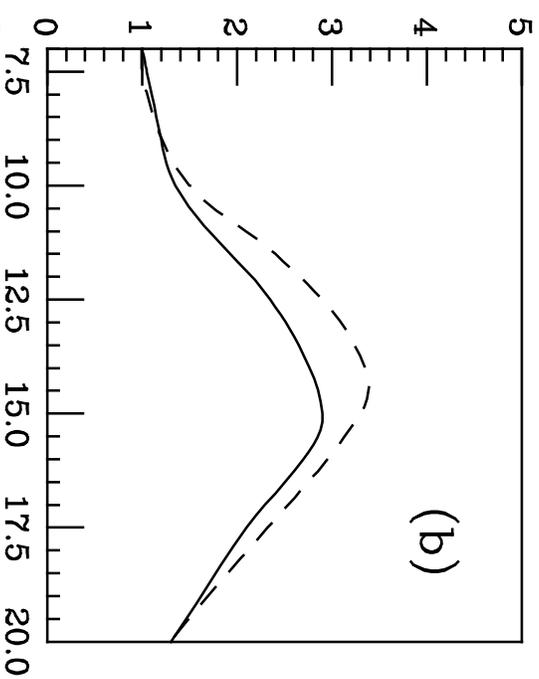
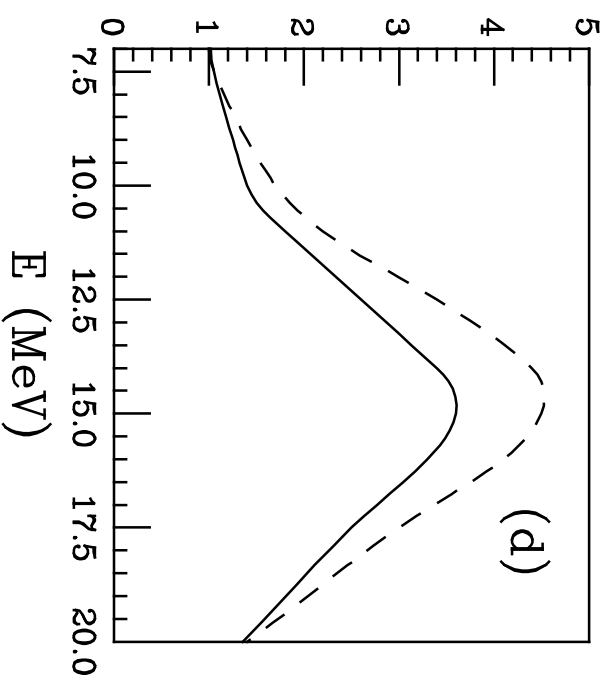
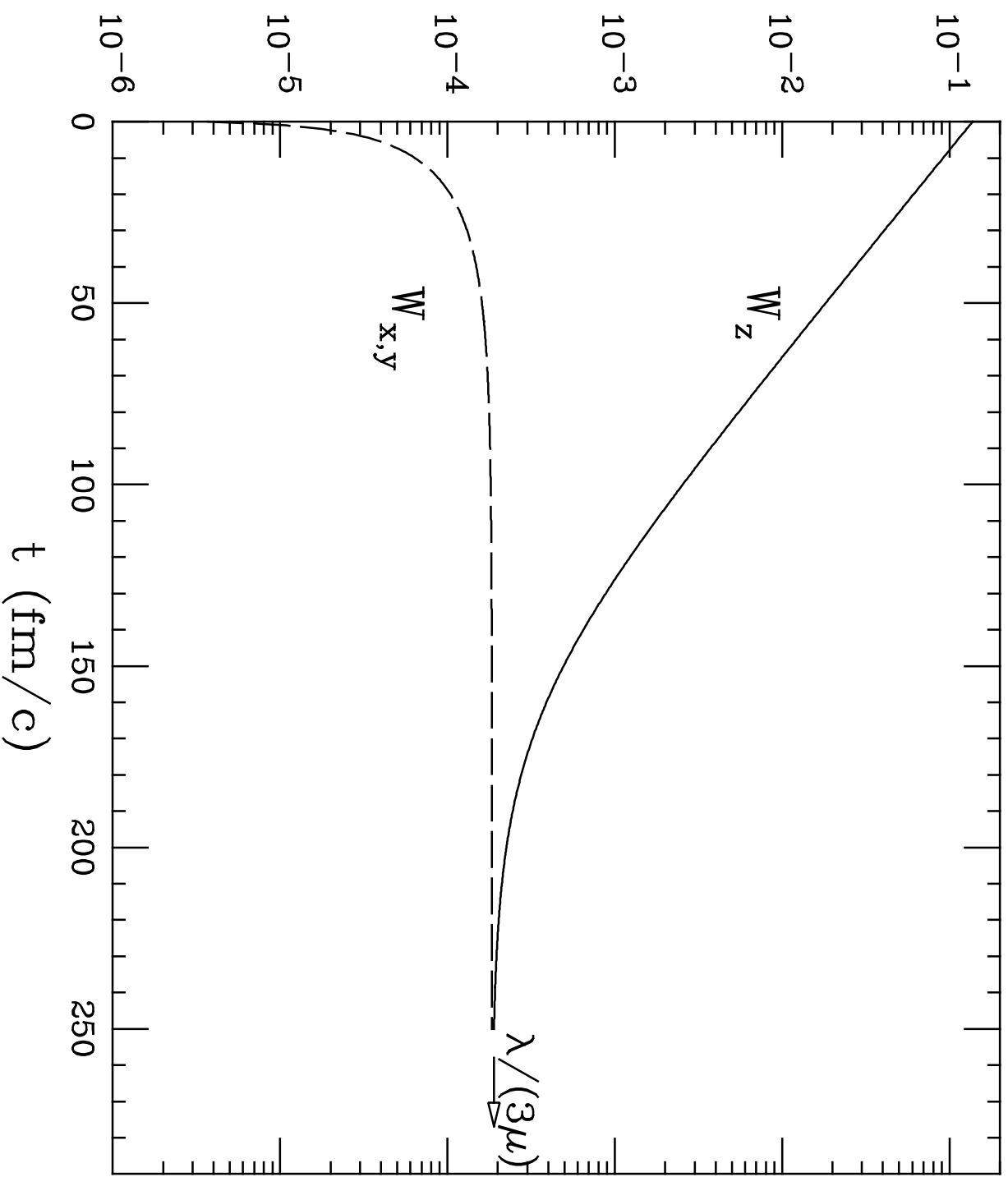
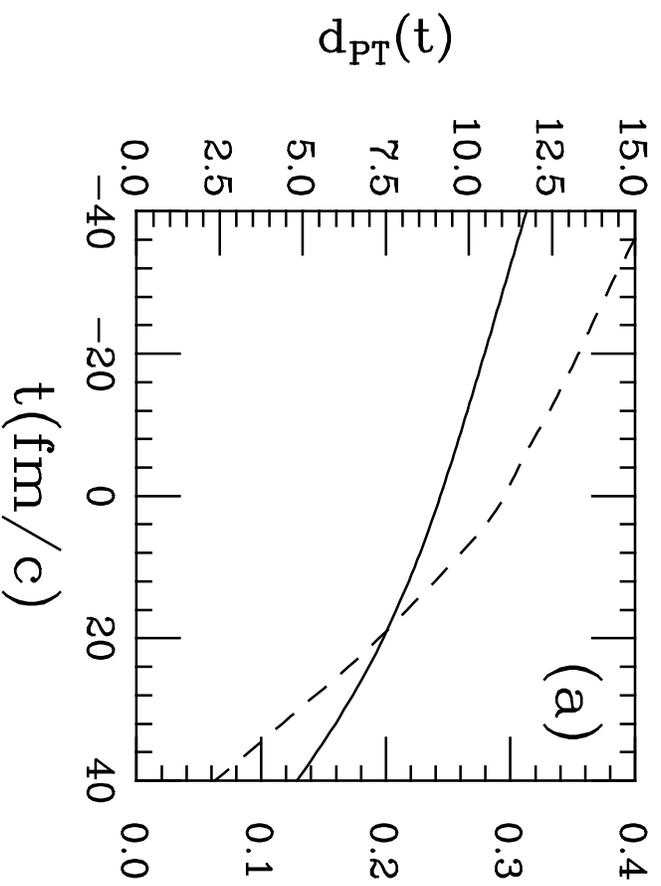
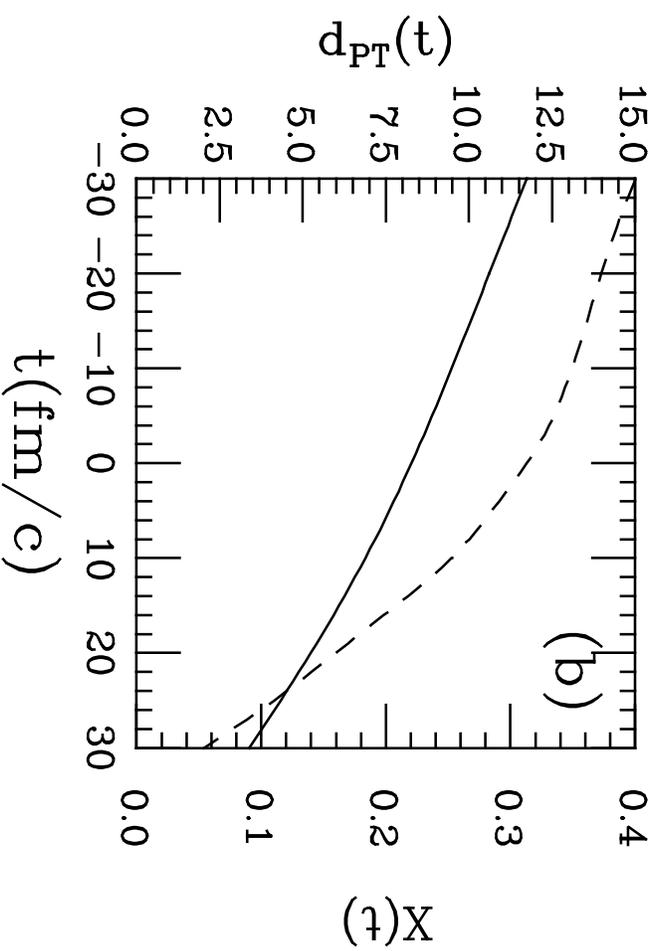
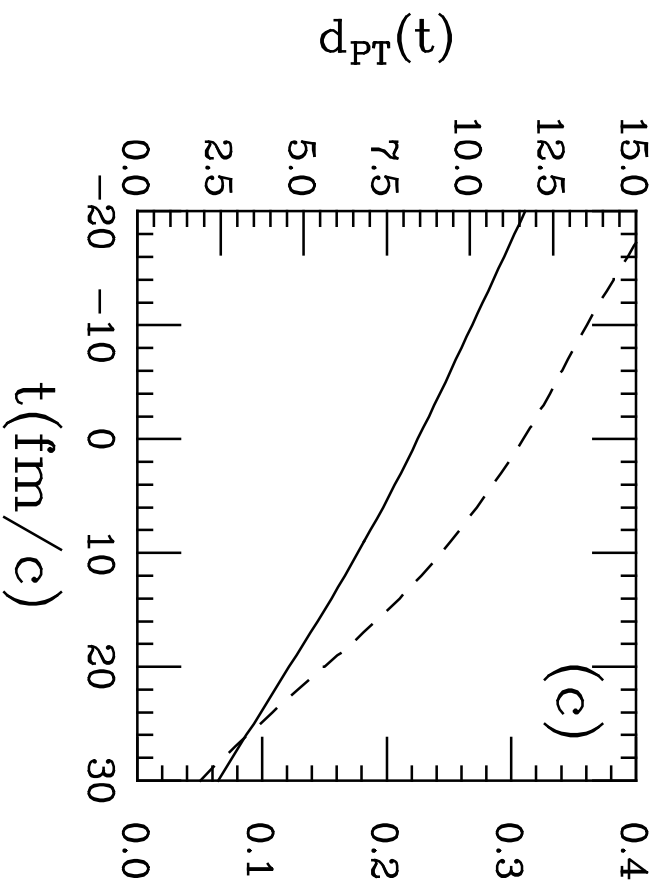
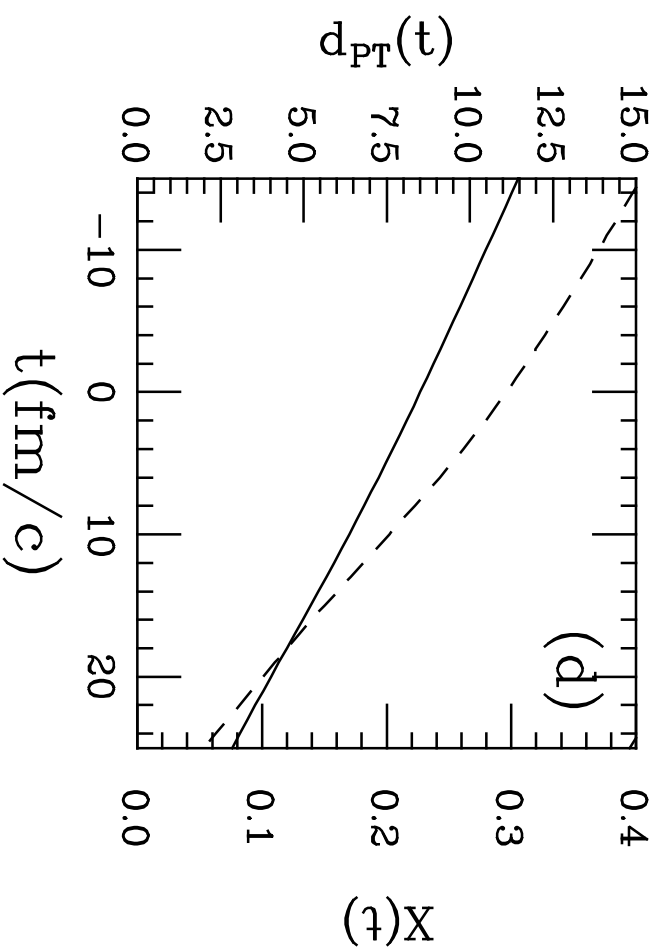
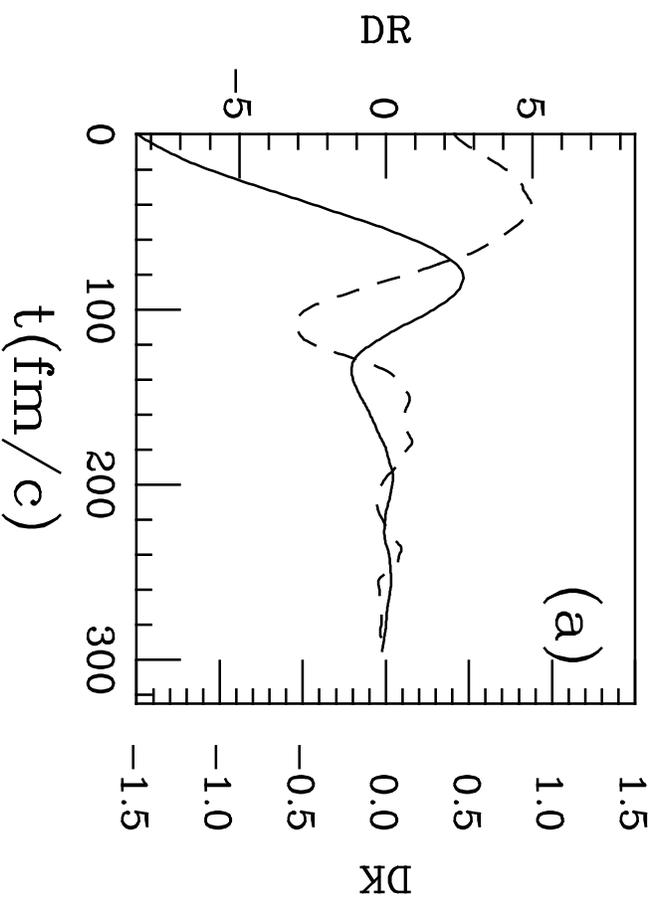
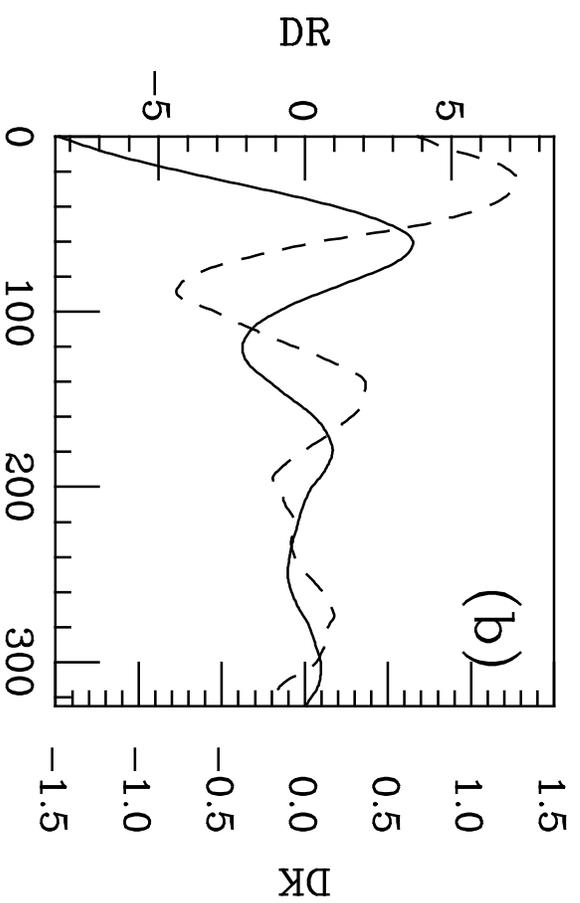
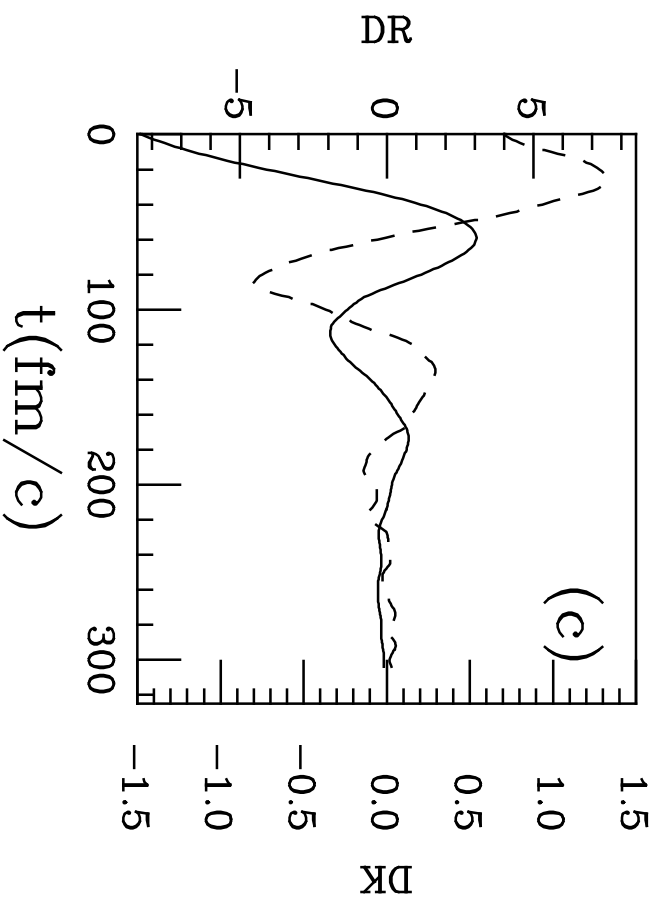
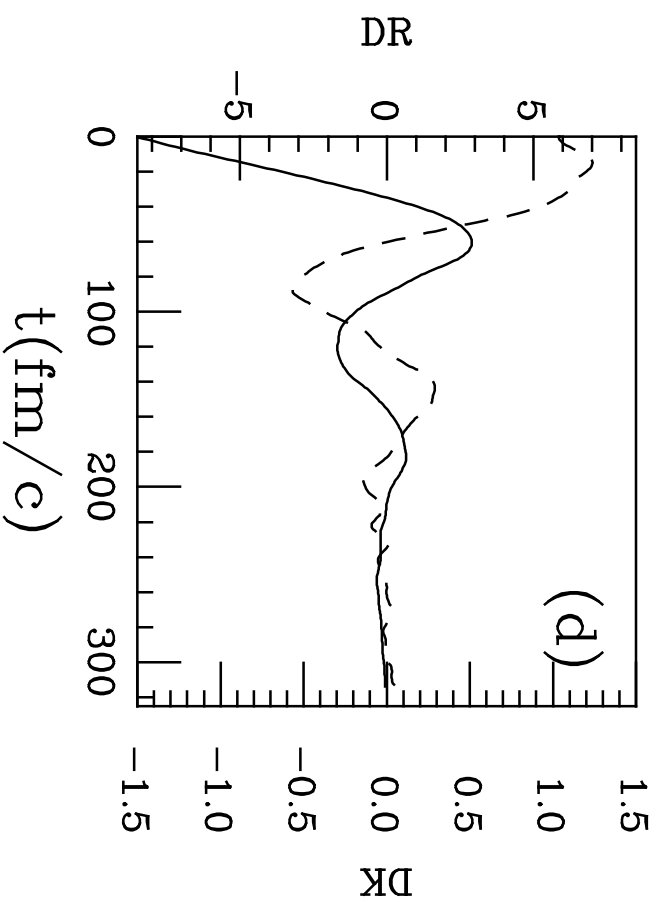
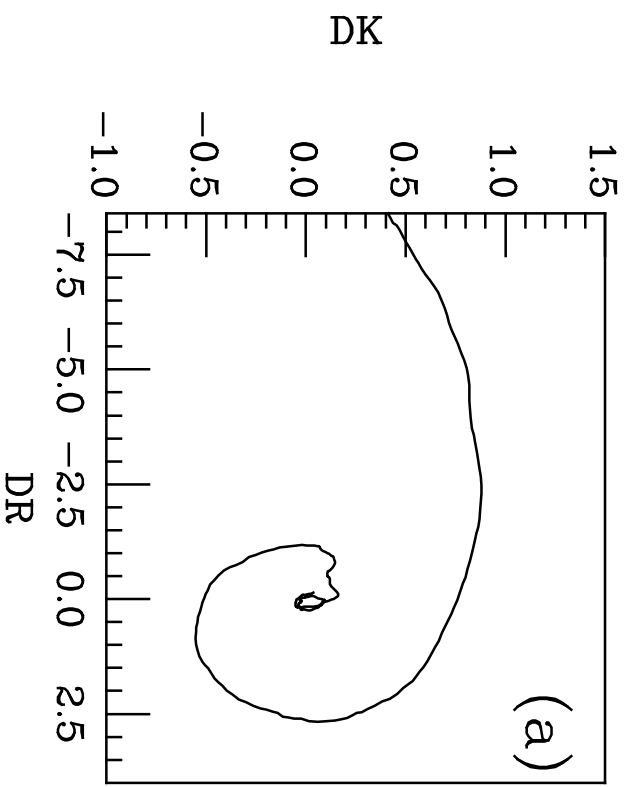
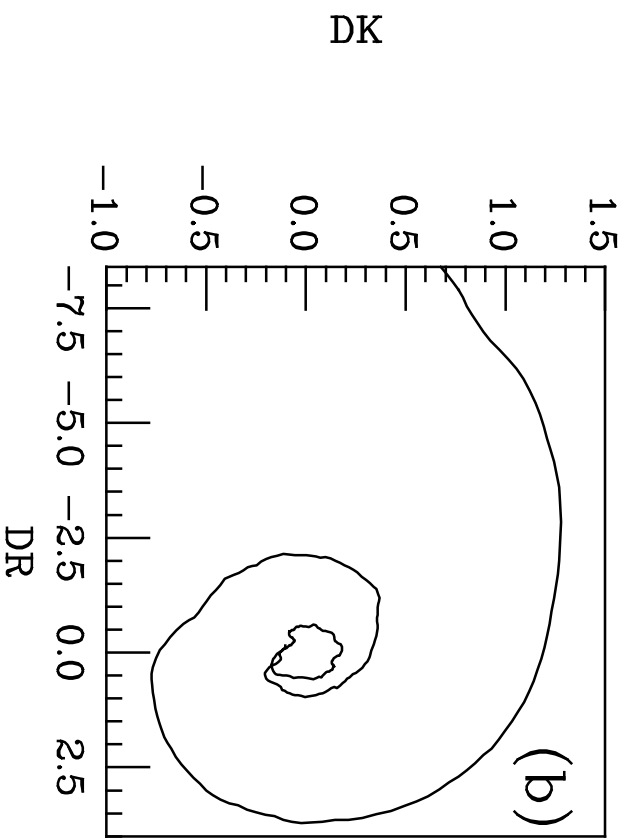
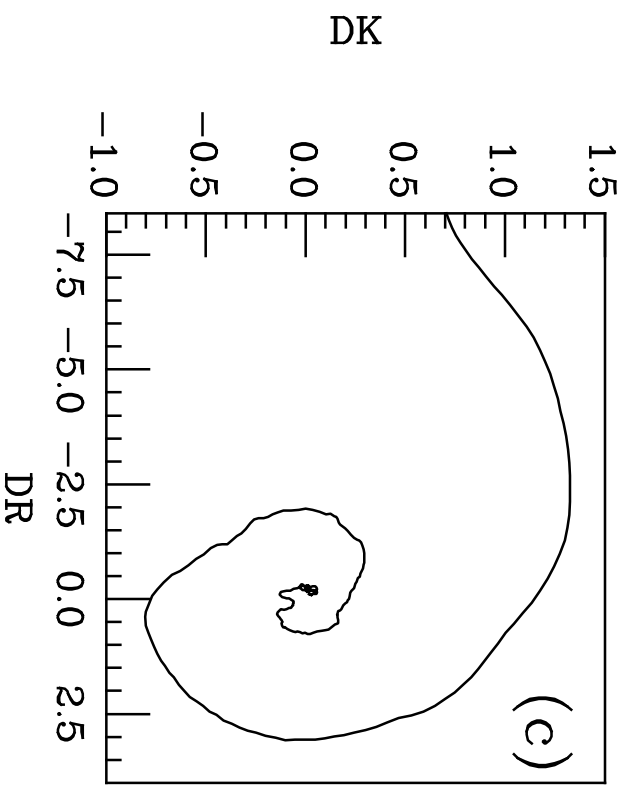
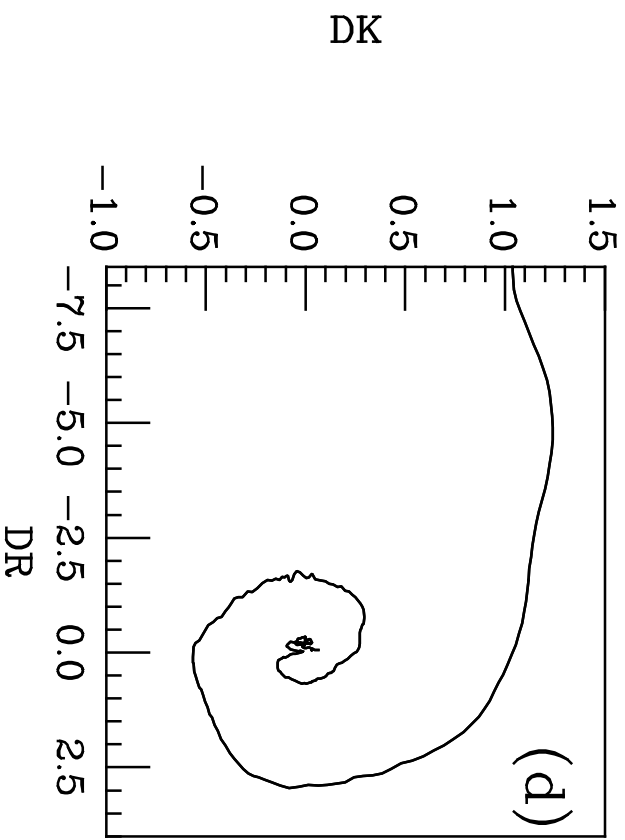
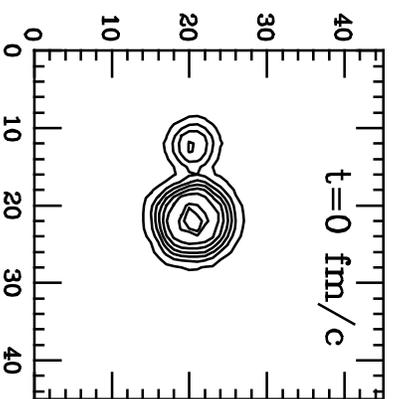
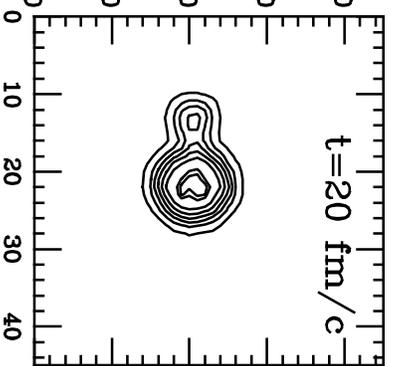
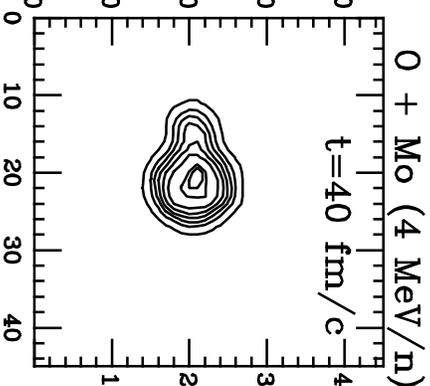
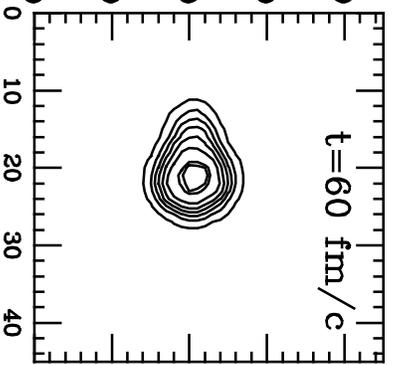
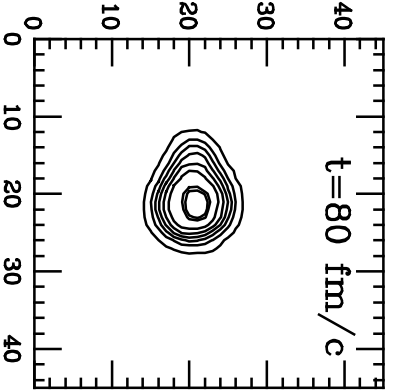
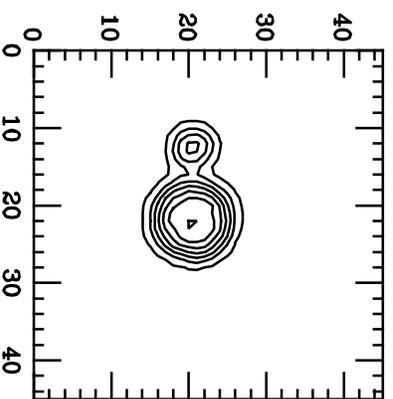
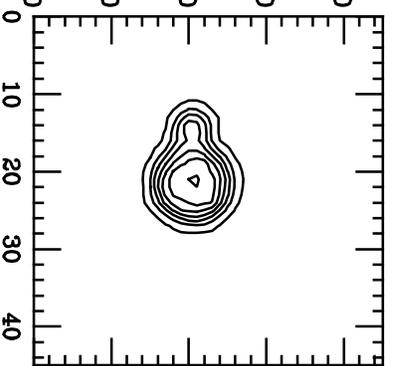
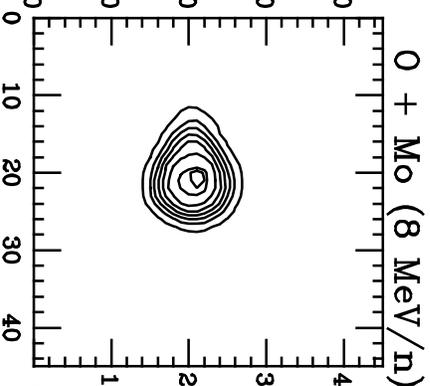
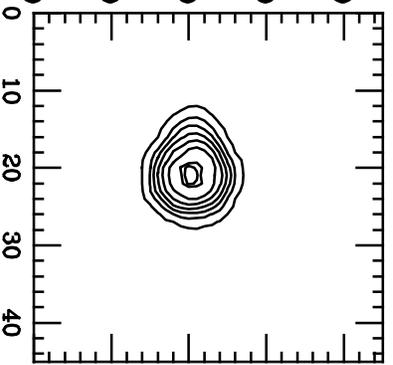
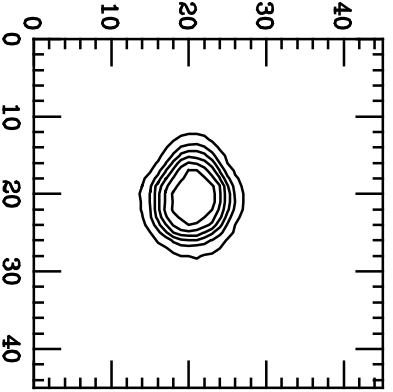
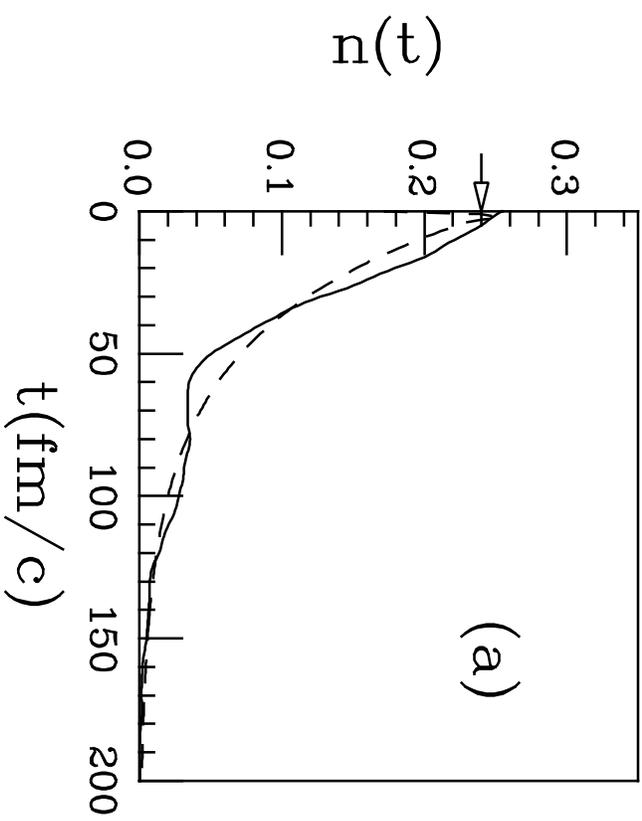
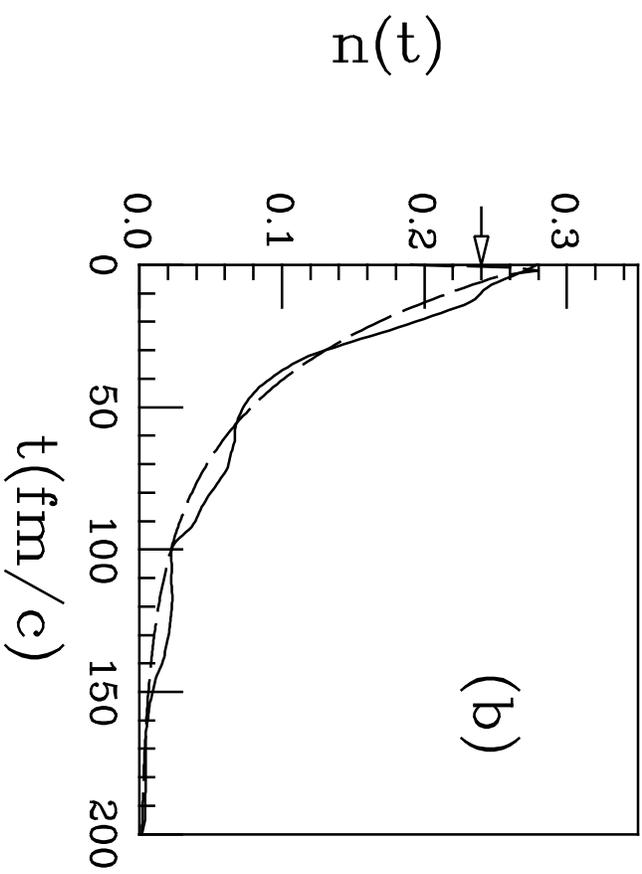
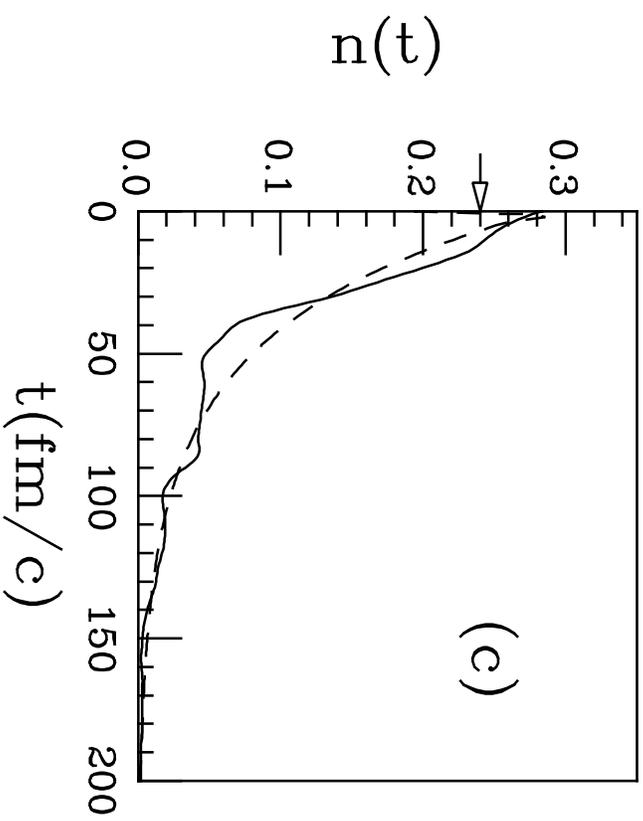
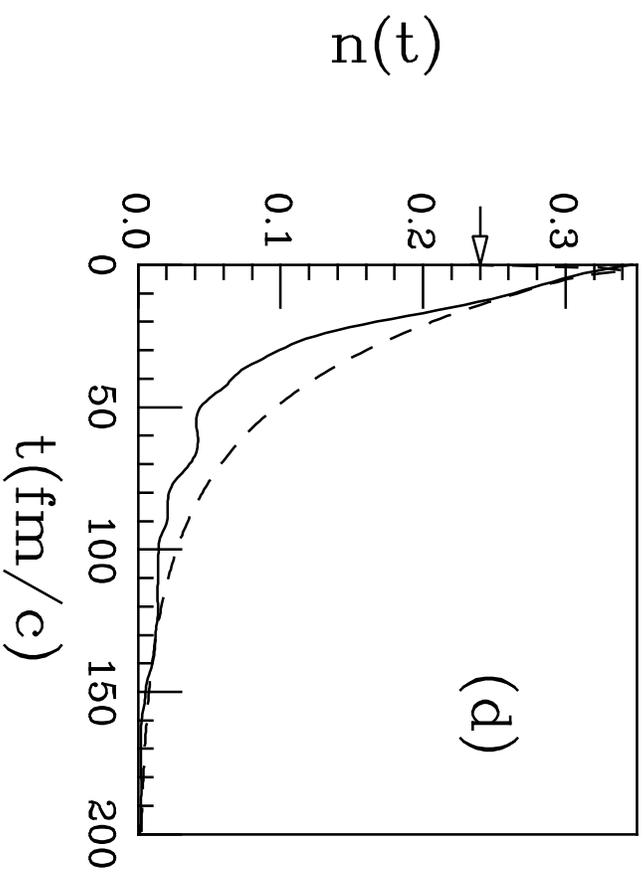
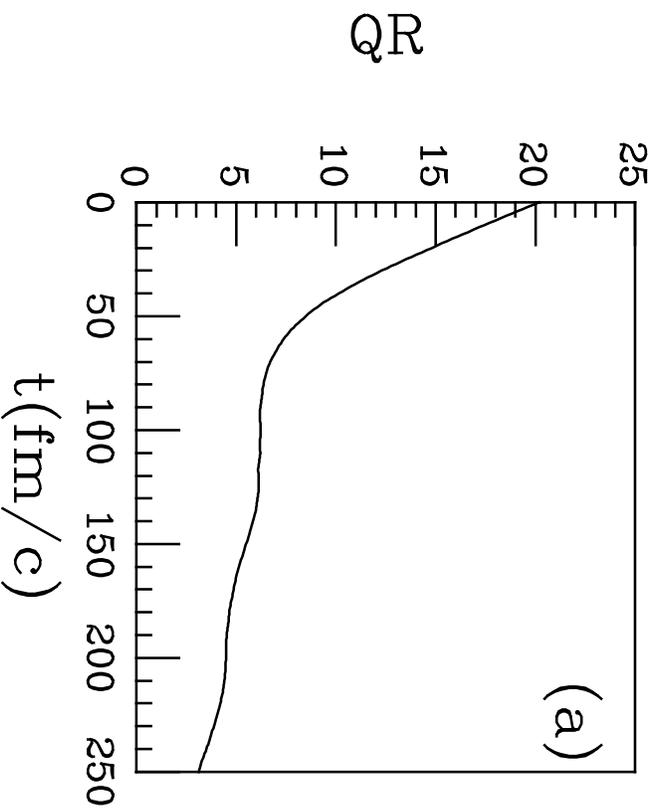
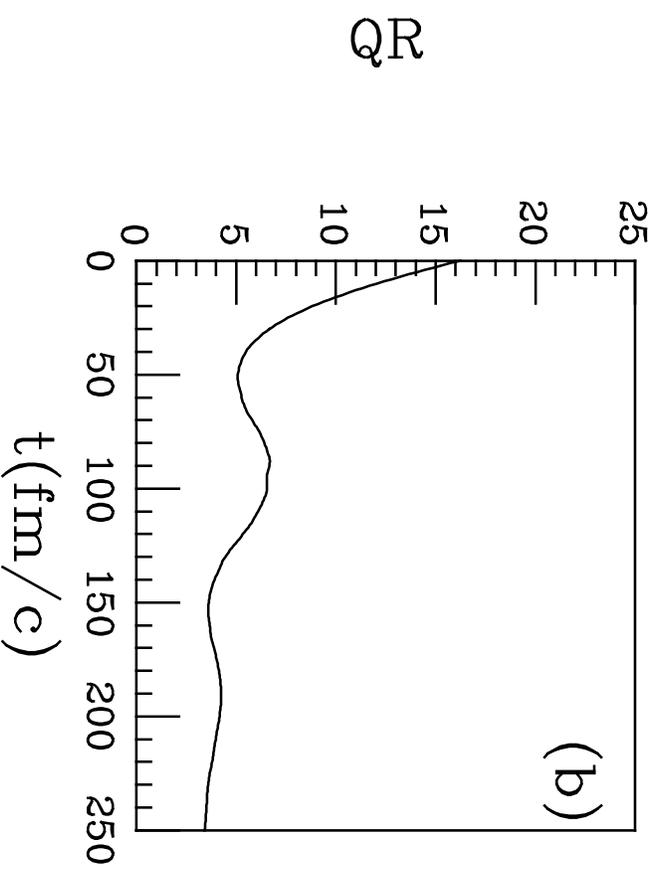
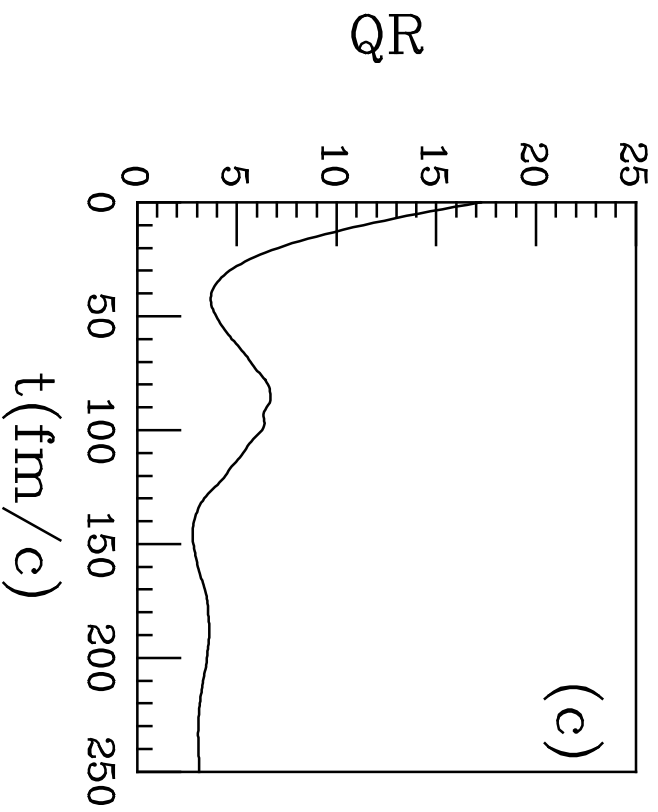
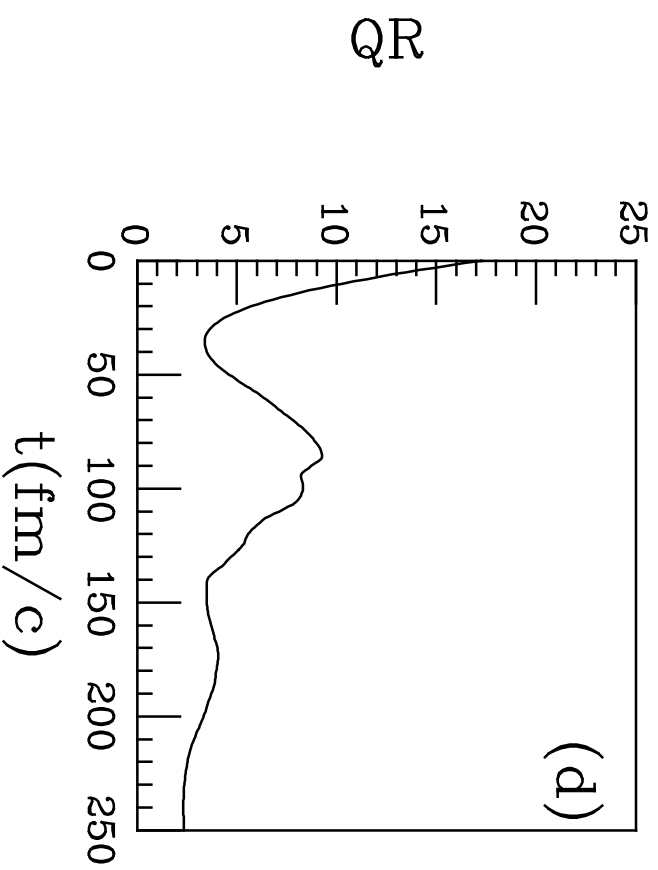
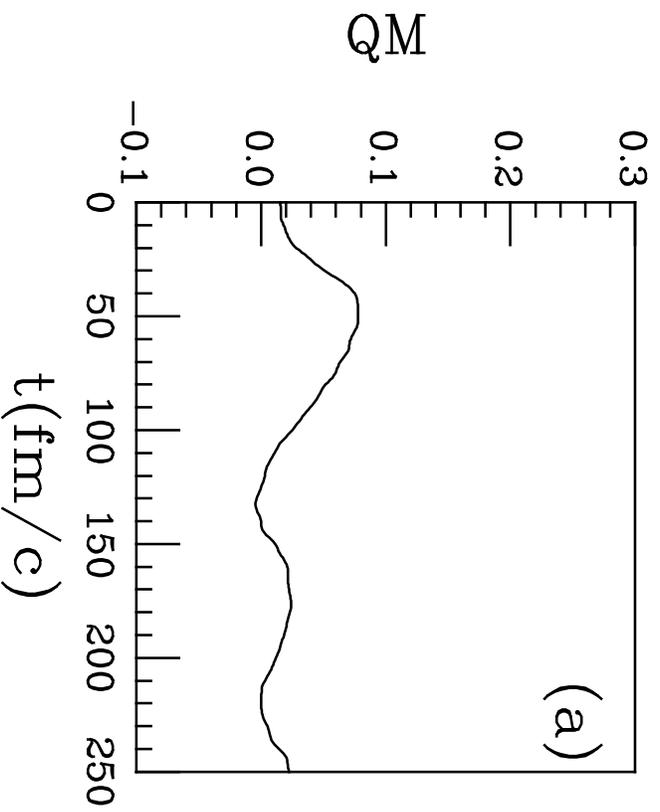
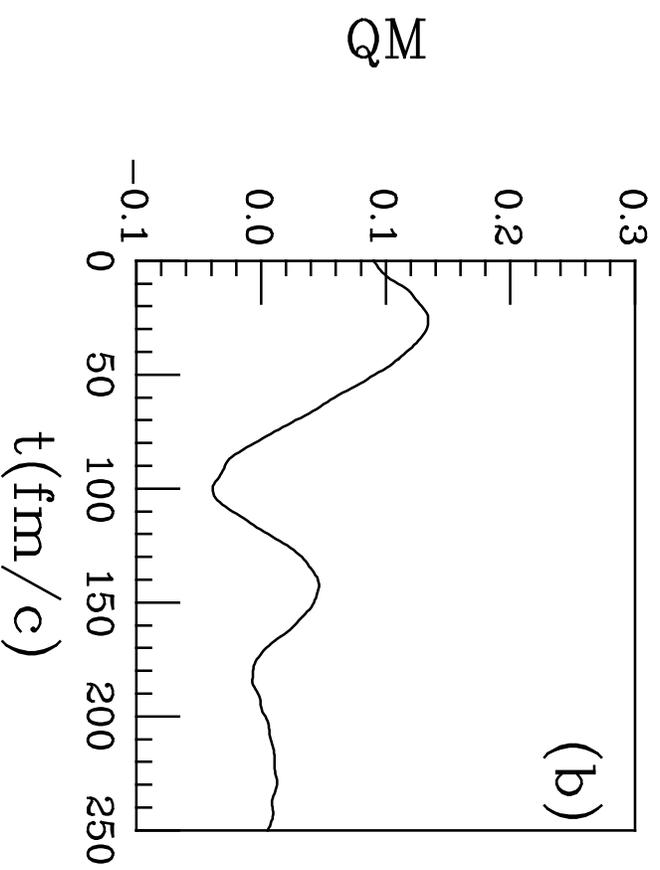
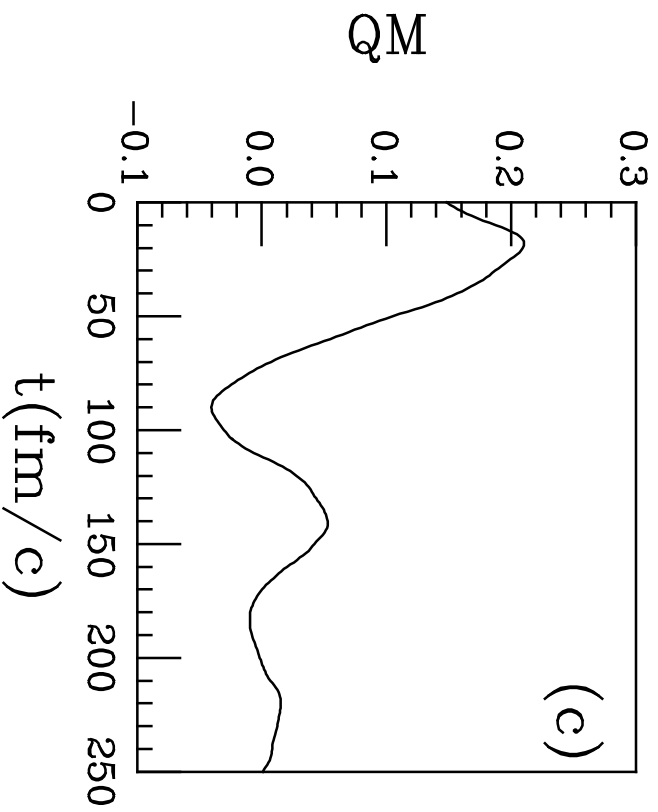
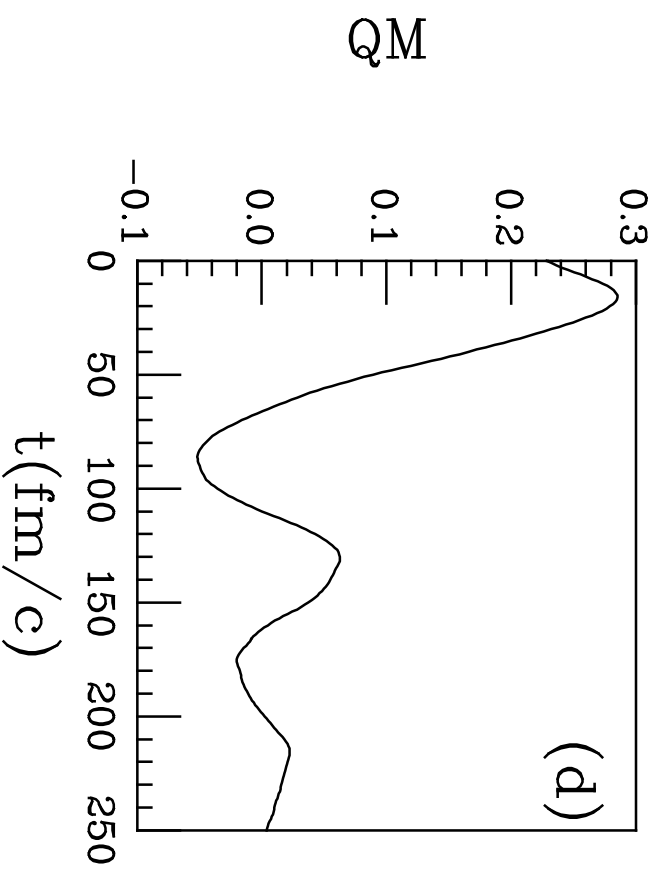
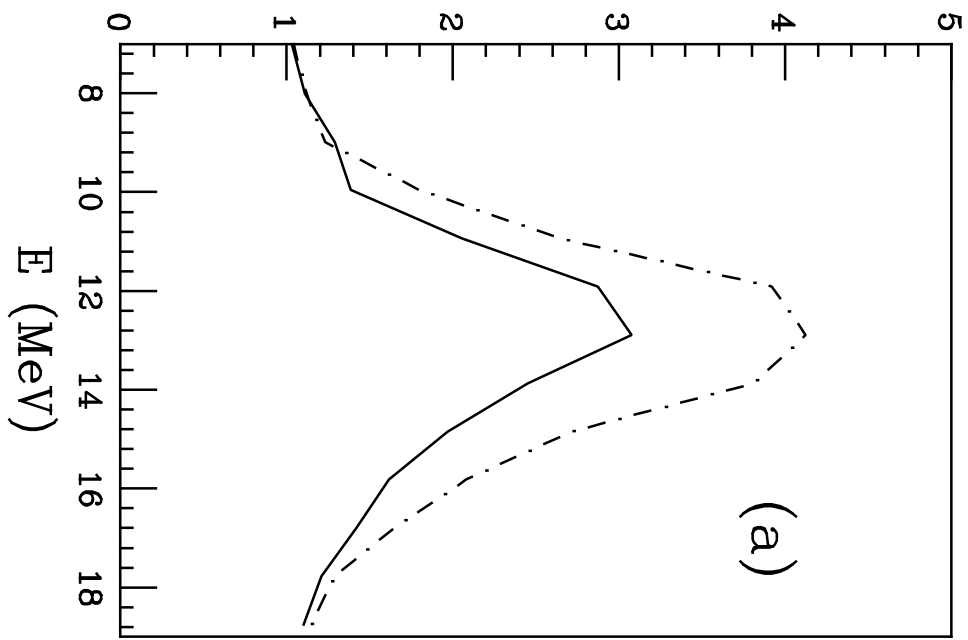
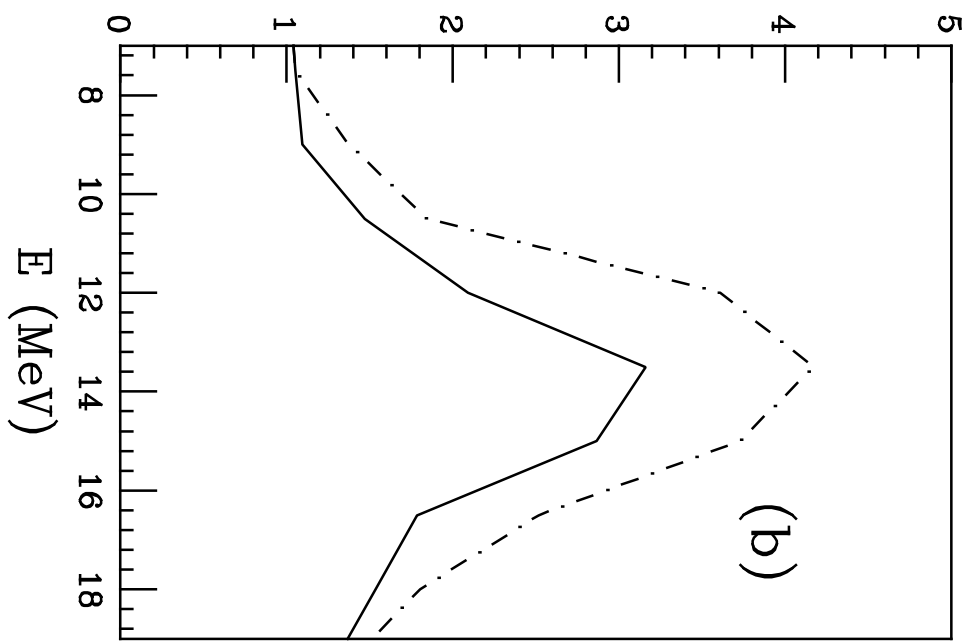
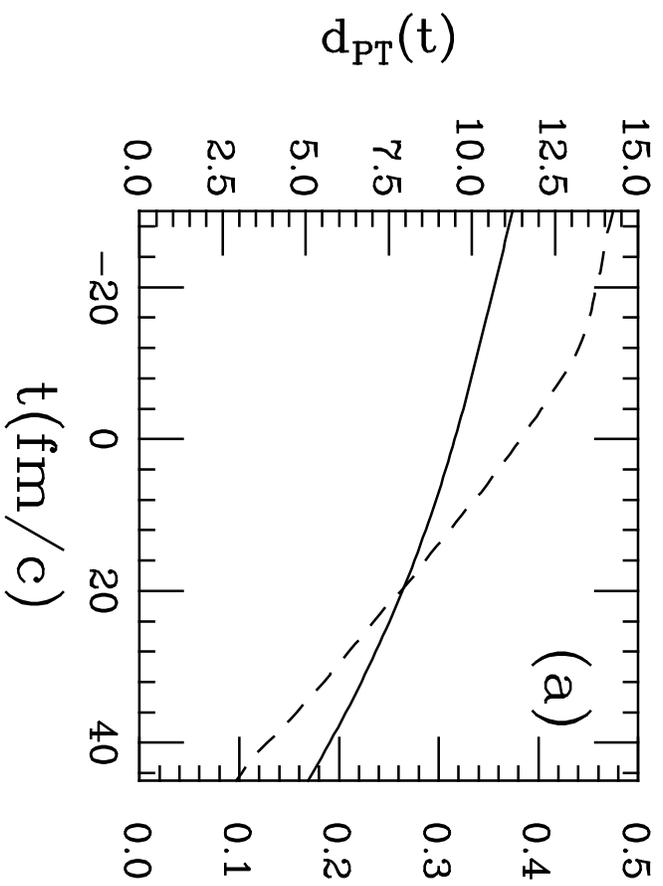
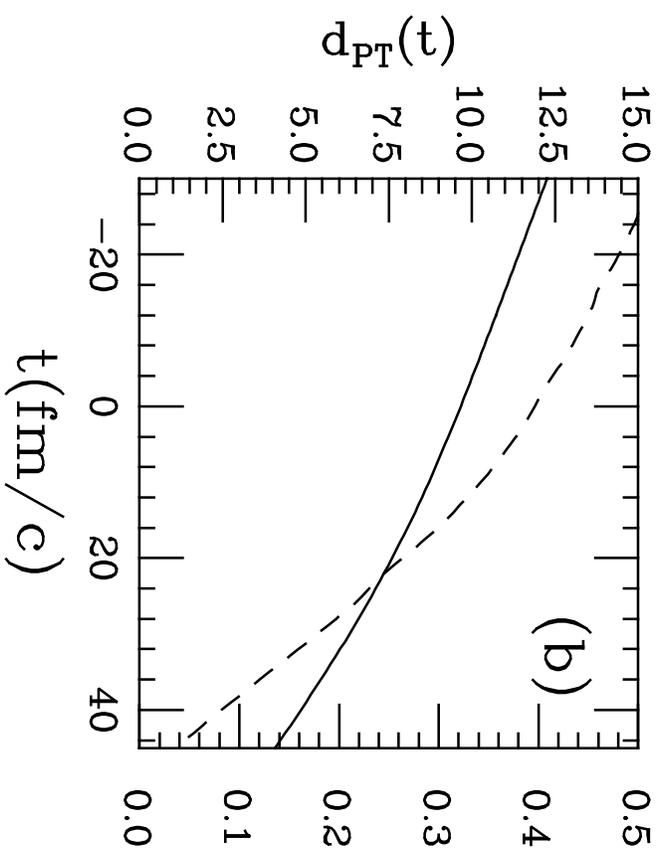
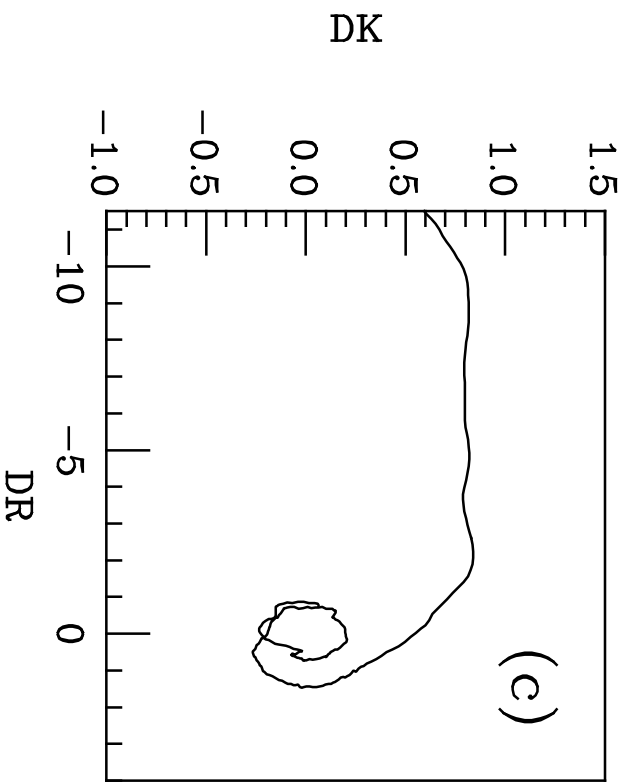
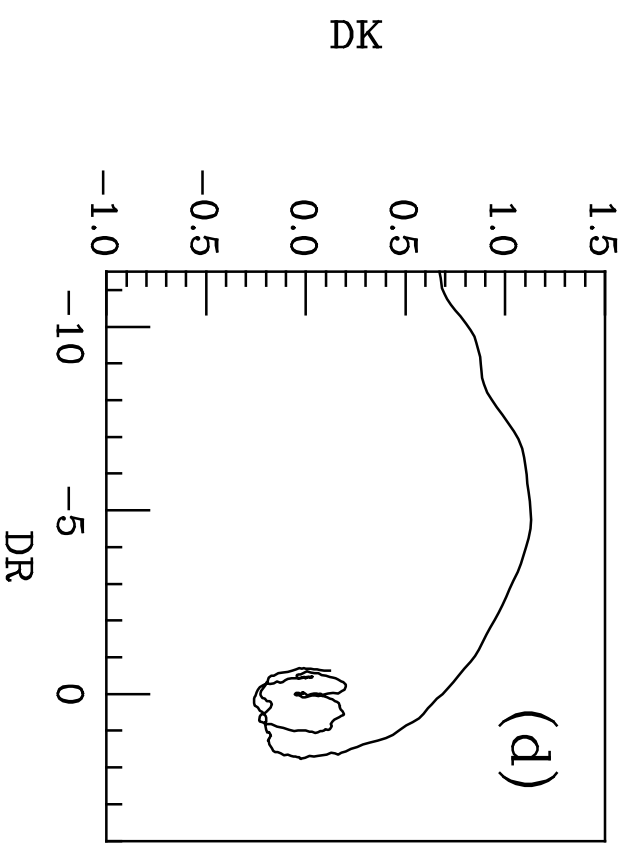
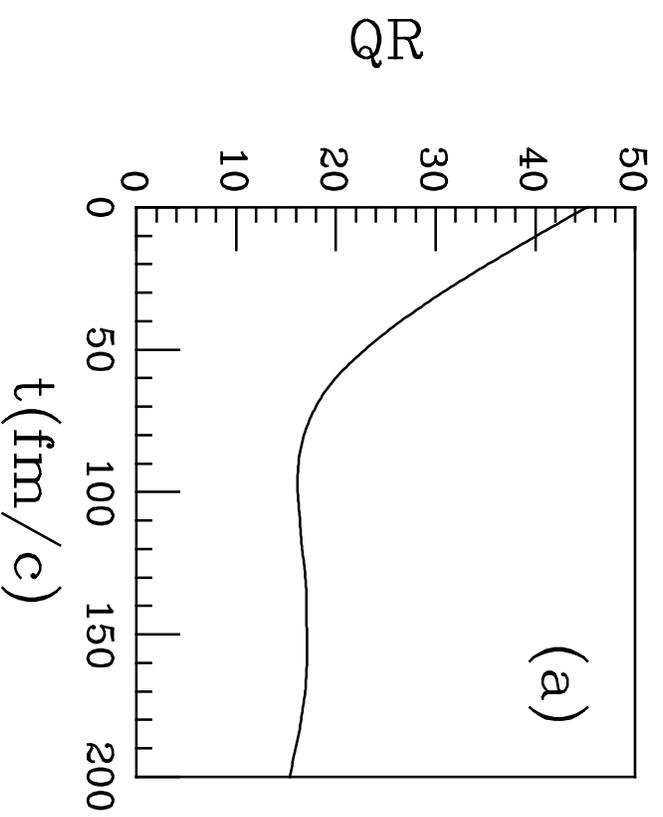
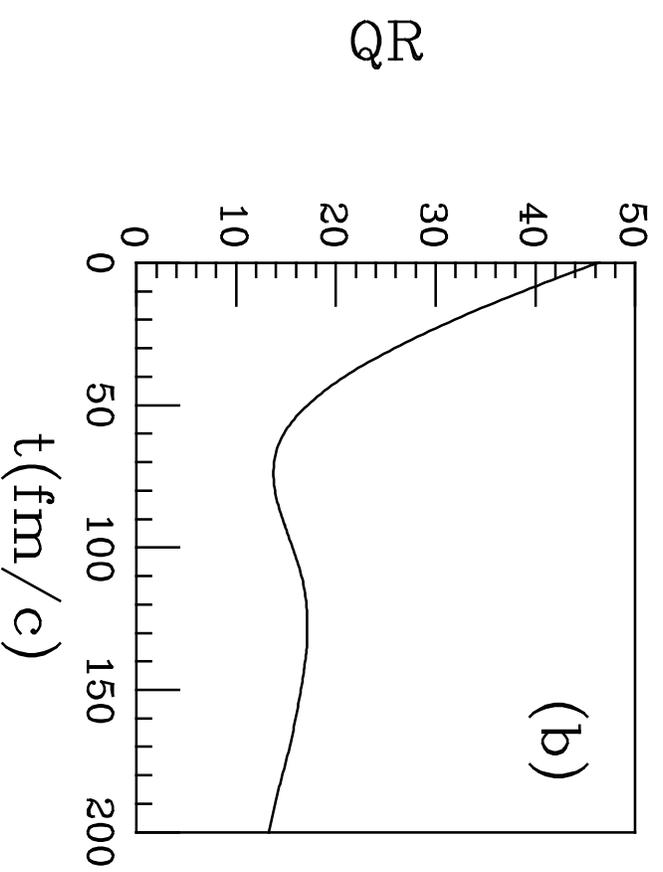
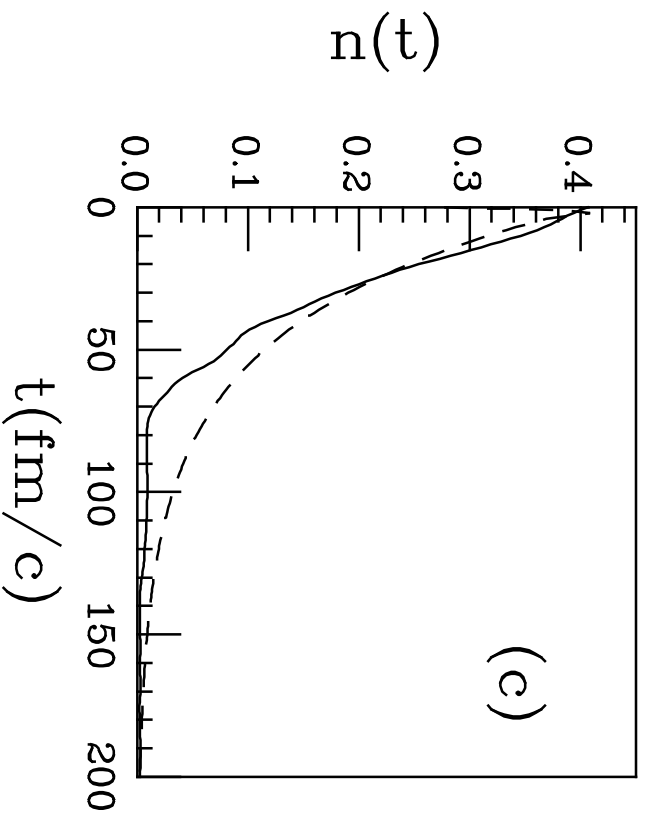
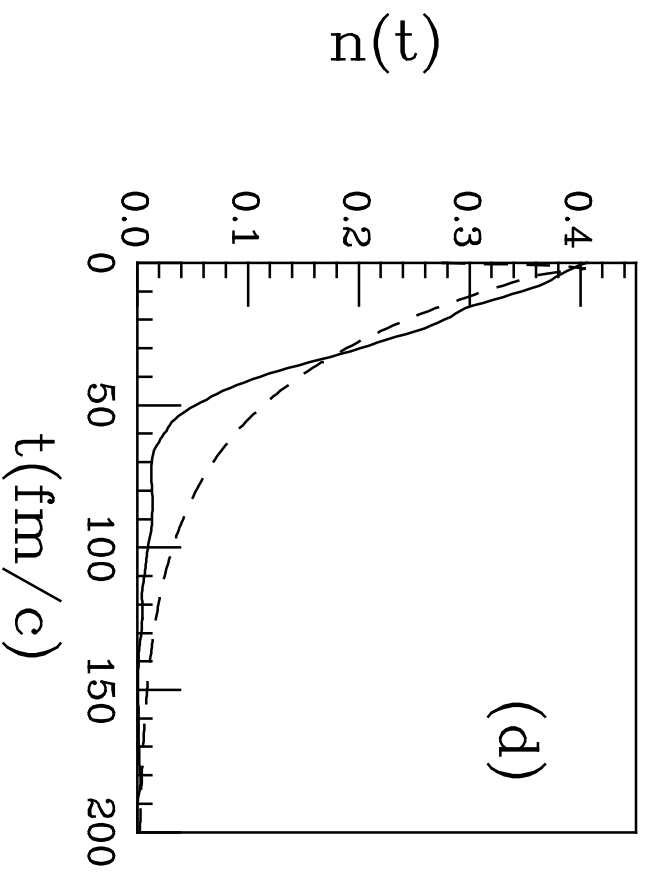
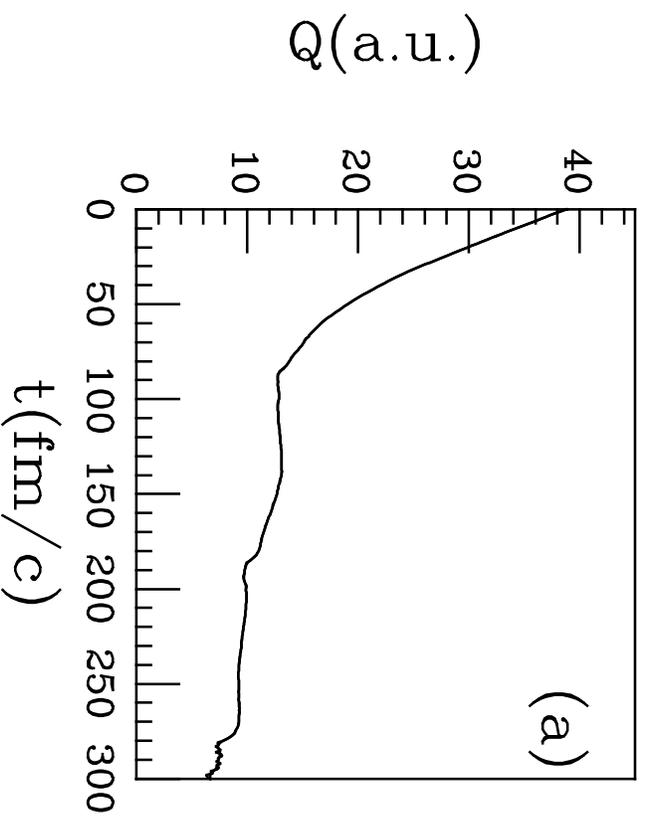
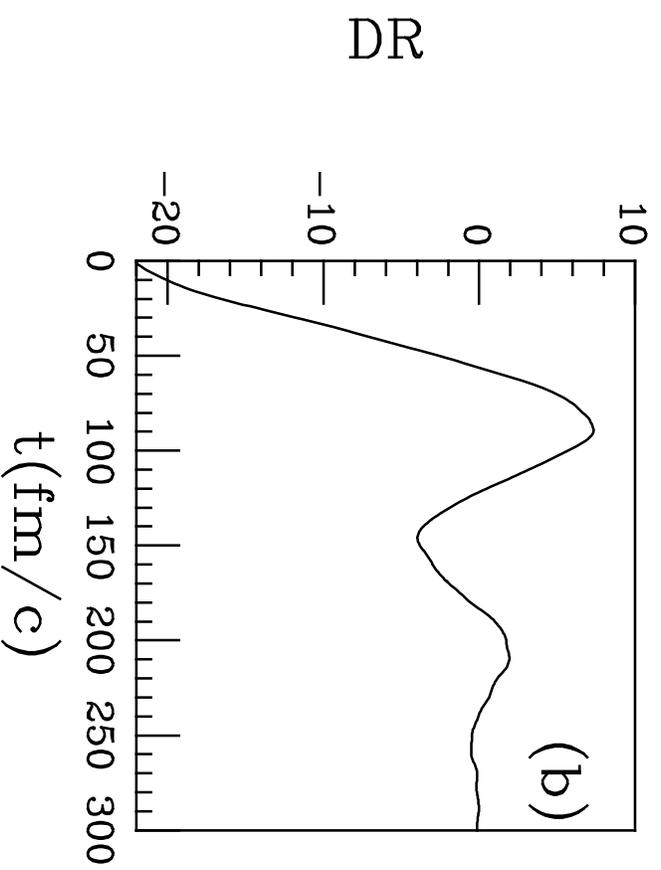
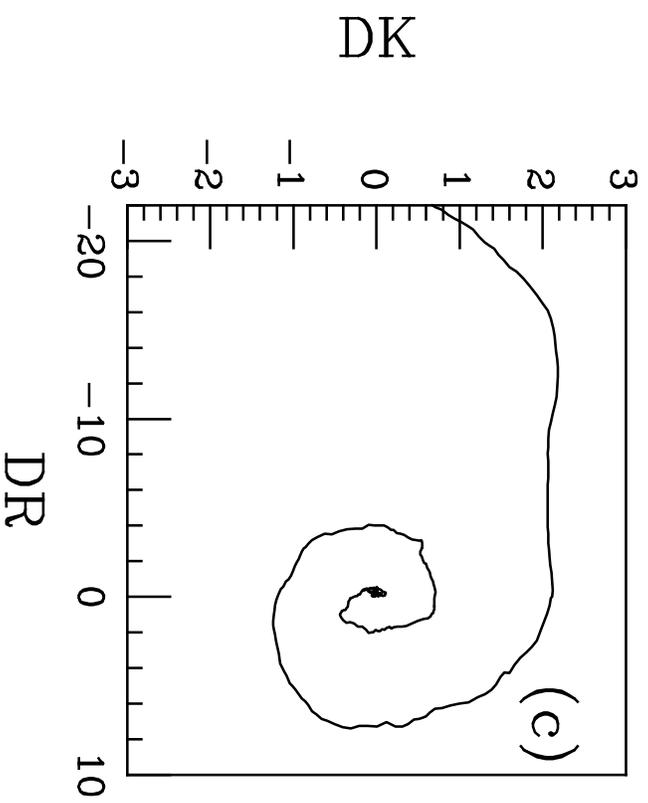
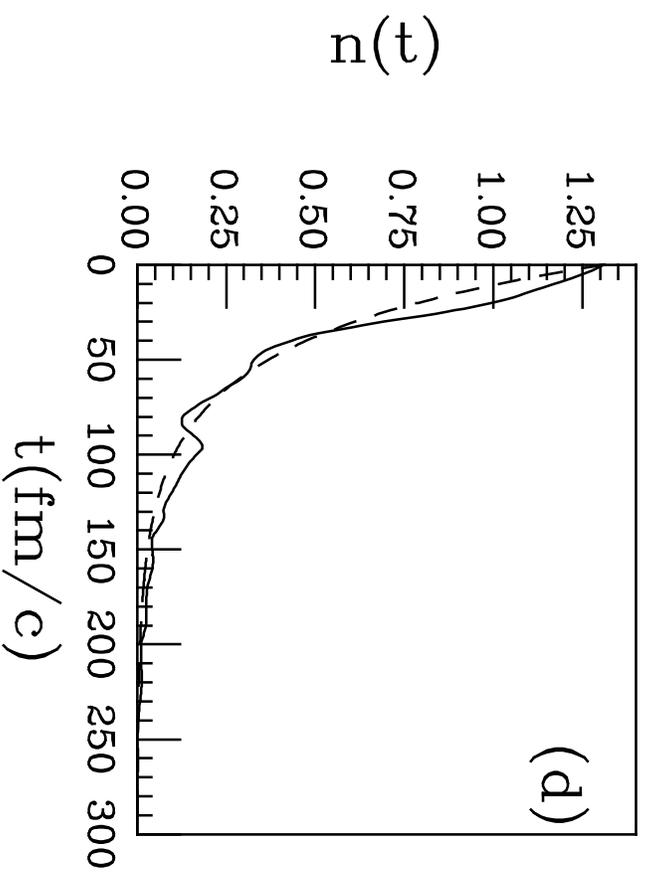
\end{document}